\documentclass[12pt,preprint]{aastex}
\usepackage{epsf}
\usepackage{amsmath, amssymb}

\shortauthors{T.Hosokawa \& S.Inutsuka}
\shorttitle{Generality of Triggered Star Formation}

\begin{document}

\title{Dynamical Expansion of Ionization and Dissociation Front around
a Massive Star. 
II. On the Generality of Triggered Star Formation}
\author{Takashi Hosokawa\altaffilmark{1} and 
     Shu-ichiro Inutsuka\altaffilmark{2}}

\altaffiltext{1}{Division of Theoretical Astrophysics, 
National Astronomical Observatory, Mitaka, Tokyo 181-8588, Japan ; 
hosokawa@th.nao.kyoto-u.ac.jp}
\altaffiltext{2}{Department of Physics, Kyoto University, Kyoto 606-8502 ;
inutsuka@tap.scphys.kyoto-u.ac.jp} 

\begin{abstract}
We analyze the dynamical expansion of the H~II region,
photodissociation region, and the swept-up shell, solving
the UV- and FUV-radiative transfer, the thermal and chemical
processes in the time-dependent hydrodynamics code.
Following our previous paper, we investigate the time evolutions
with various ambient number densities and central stars.
Our calculations show that basic evolution is qualitatively 
similar among our models with different parameters.
The molecular gas is finally accumulated in the shell, and the 
gravitational fragmentation of the shell is generally expected.
The quantitative differences among models are well understood
with analytic scaling relations.
The detailed physical and chemical structure of the shell is
mainly determined by the incident FUV flux and the column density
of the shell, which also follow the scaling relations.
The time of shell-fragmentation, and the mass of the 
gathered molecular gas are sensitive to
the ambient number density. 
In the case of a low density,
the shell-fragmentation occurs over a longer timescale,
and the accumulated molecular gas is more massive than in the
case of a high density. 
The variations with different central stars are more moderate.
The time of the shell-fragmentation differs by a factor of several
with the various stars of $M_* = 12-101~M_\odot$.
According to our numerical results, we conclude that the expanding
H~II region should be an efficient trigger for star formation
in molecular clouds if the mass of the ambient molecular
material is large enough. 
\end{abstract}

%%%%%%%%%%%%%%%%%%%%%%%%
%%%%%%%%%%%%%%%%%%%%%%%%
\section{Introduction}
%%%%%%%%%%%%%%%%%%%%%%%%
%%%%%%%%%%%%%%%%%%%%%%%%

Massive stars are ubiquitous in many star-forming regions, and
their feedback processes have  
significant impacts on the surrounding interstellar medium (ISM). 
UV/FUV radiation photoionizes/photodissociates molecular gas,  
and the H~II and photodissociation regions (PDR) 
expand around the massive star. 
The stellar winds and supernovae explosions give rise to  
expanding bubbles in ISM. 
The role of these feedback processes in relation to ISM has 
been studied from two competing perspectives on their effects
on star-formation activity.
The expanding H~II region and bubble erode the parental molecular clouds
and suppress star formation ({\it negative feedback}).
On the other hand, the shock front (SF) emerges as the high-pressure
region expands. The SF compresses the ISM and star 
formation can be triggered in very dense layers 
({\it positive feedback}).

In this paper, we focus on the H~II regions and
their feedback processes. 
The role of the expanding H~II regions is also discussed in 
terms of negative and positive feedback processes.
Most of the massive stars are born in the giant molecular clouds (GMCs) 
and H~II regions begin to expand in the dense cores and clumps in GMCs 
\citep[ultra compact H~II regions, e.g.,][]{Cw02}.
As the H~II region expands, the molecular gas is gradually eroded. 
When the IF reaches the cloud surface, the ionized gas spouts from
the cloud (``champagne flow'', see Tenorio-Tagle 1979).
The molecular material around massive stars is dispersed,
so that star formation is quenched \citep[e.g., ][]{Wt79,FST94}.
On the other hand, the SF emerges in front of the IF owing to
the pressure excess of the H~II region. 
The SF sweeps up the ISM to form a shell and star formation 
will be induced by the instability of the shell 
\citep[``collect and collapse'' scenario ; see, e.g.,]
[and references therein]{EL77, Elm89}.
The SF also compresses pre-existing globules through their 
photoevaporation and star formation will be triggered 
\citep[radiative implosion ; see, e.g.,][]{Bt89, BM90}.

The hydrodynamical expansion of the H~II regions has been 
studied numerically \citep[e.g.,][]{Mt65, Tt79, FTB90, GF96}.
These efforts have successfully shown the various
dynamical aspects of the expansion in both homogeneous and
inhomogeneous ambient media. However, these do not include
the outer PDR and thermal processes that should be dominant there.
\citet{HH78} have analytically shown that the dissociation
fronts (DFs) initially expand outside the SF, but finally
slow down to be taken into the shell.
\citet{RD92} and \citet{DFS98} studied the expansion of the H~II region
and outer PDR around a single massive star, solving the radiative 
transfer of UV and FUV photons. 
Although the hydrodynamics has not been included,
they have shown that a significant amount of the molecular gas
is not ionized but photodissociated. \citet{DFS98}
have pointed out that the photodissociation of the molecular gas
reduces the star-forming capacity of the cloud

\citet{Dv03} have provided a good simple example 
to investigate the role of the expanding H~II region. 
They have observed the molecular fragments
around the galactic H~II region, Sharpless 104 (Sh104), and found a 
young cluster within the core of one fragment. 
They have argued that this is 
clear evidence of the ``collect and collapse'' model. 
In our previous paper \citep[][hereafter, Paper I]{HI05}, 
we have modeled Sh104 performing the hydrodynamical
calculation including the PDR and the swept-up shell
as well as the H~II region.
Our calculation has shown that the molecular shell
forms around the H~II region shielding the FUV radiation.
The abundance of the accumulated molecules shows excellent agreement
with the observation.
We have also suggested a role for the expanding H~II regions as a positive
feedback process. The swept-up mass remains in the shell and the shell
can be dominated by the cold molecular gas shielding UV and FUV
photons. 

Recently, \citet{Dv05} have reported dozens of galactic H~II regions
similar to Sh104, which show that such triggering is a common
process in some situations.
So, the next interesting questions to answer are the following :
\begin{itemize}
\item
How does the time evolution change with differing ambient 
number density and mass of the central star? 
Is the molecular gas always accumulated in the shell? 
If the evolution change with a different ambient density
or central star, why so?
\end{itemize}
To answer these questions, we investigate the time evolution
of the H~II region, PDR, and the swept-up shell with various
ambient number densities and central stars.
As in Paper I, the UV- and FUV-radiative transfer
and the thermal and chemical processes are solved 
using a time-dependent hydrodynamics code.
We show the numerical results of some representative models
and explain what causes the differences among models.
We can show that, in many cases, the FUV photons are shielded 
and the molecular gas is finally gathered in the shell.

The organization of this paper is as follows: 
In \S2, we explain the input physics of our code. 
The subsequent \S3 is devoted to showing the procedures of
the numerical calculation. 
\S4 and 5 are the main part of this paper, where the numerical
results are shown. The dependence on the ambient number density,
and on the luminosity (mass) of the central star is studied
in each section. \S6 and 7 are assigned to discussions and 
conclusions.

%%%%%%%%%%%%%%%%%%%%%%%%%%%%%%%%%%%%%%%%
%%%%%%%%%%%%%%%%%%%%%%%%%%%%%%%%%%%%%%%%
\section{Input Physics for Calculation}
%%%%%%%%%%%%%%%%%%%%%%%%%%%%%%%%%%%%%%%%
%%%%%%%%%%%%%%%%%%%%%%%%%%%%%%%%%%%%%%%%

%%%%%%%%%%%%%%%%%%%%%%%%%%%%
\subsection{Basic Equations}
\label{ssec:beq}
%%%%%%%%%%%%%%%%%%%%%%%%%%%%

%-------------------------------------%
\subsubsection{Hydrodynamic Equations}
%-------------------------------------%

In this paper, we consider the spherical expansion of the H~II region. 
The hydrodynamic equations of the system are written as,
\begin{equation}
\label{eq:rzku}
\frac{d}{dt} \left( \frac1\rho \right)
- \frac{\partial (r^2 u)}{\partial m} = 0 \ ,
\end{equation}
\begin{equation}
\label{eq:mmtm}
\frac{du}{dt} + r^2 \frac{\partial p}{\partial m} = 0 \ ,
\end{equation}
\begin{equation}
\label{eq:egyeq}
\rho \left( 
\frac{d E}{d t} + \frac{\partial (r^2 u p)}{\partial m}
+ \frac{\partial (r^2 q_{\rm cond})}{\partial m} \right)
= n ( \Gamma - \Lambda ) \ ,
\end{equation}
\begin{equation}
\label{eq:eos}
p = 
\left\{ 0.5 (1 + 3 X_{{\rm H}^+} + X_{\rm H} ) + Y_{\rm He} \right\} 
n k T \ ,
\end{equation}
\begin{equation}
\label{eq:rhon}
\rho = (1 + 4 Y_{\rm He}) m_{\rm H} n \equiv \mu m_{\rm H} n \ ,
\end{equation}
where $m$ is the mass coordinate defined as $dm = \rho r^2 dr$,
$\rho$ is the mass density, $n$ is the number density of the
hydrogen nucleon, $u$ is the velocity, $p$ is the gas pressure, and
$E$ is the total energy, including the kinetic energy ($u^2/2$) and 
the thermal energy ($e \equiv p/(\gamma - 1)/\rho$) per unit mass. 
We adopt $\gamma = 5/3$ for simplicity.
In the above equations, $X_{{\rm H}^+}$ $(X_{\rm H})$ is the number ratio
defined as $n_{{\rm H}^+}/n$ $(n_{\rm H}/n)$, and $Y_{\rm He}$ is the 
abundance of He atoms.
In the energy equation, $\Gamma$ and $\Lambda$ represent the heating and 
cooling rates per hydrogen nucleon, and $q_{\rm cond}$ is the heat
transfer by thermal conduction.
We include 5 heating processes and 12 cooling processes to calculate
$\Gamma$ and $\Lambda$ (see \S\ref{ssec:thm}). 
The thermal conduction is needed to satisfy the ``Field condition''
(see Appendix \ref{sec:field}).
We do not include the self-gravity of the gas to 
study the simple situation of the HII region expanding into a 
homogeneous medium.
Even if we include the gas self-gravity, the H~II region quickly
expands at the velocity of $\sim 10~km/s$, and the smaller inward
velocity caused by the self-gravity does not significantly
affect our results.
 The gravity of the central star does not affect the evolution 
of H~II regions, unless the H~II region is very small.
The radius where the velocity of escape from the star is comparable to the 
sound speed of the H~II region is $\sim 10^{-3}~(M_*/10 M_\odot)~{\rm pc}$,
where $M_*$ is the stellar mass. This is much smaller
than the initial Str\"omgren radius, 
\begin{equation}
R_{\rm st} = 0.64 \ {\rm pc} 
\left( \frac{S_{\rm UV}}{10^{49} {\rm ~s}^{-1}} \right)^{1/3}
\left( \frac{T_{\rm HII}}{10^4 {\rm ~K}} \right)^{1/4}
\left( \frac{n}{10^3 {\rm ~cm}^{-3}} \right)^{-2/3} \ ,
\label{eq:rst}
\end{equation}
with the number density considered in this paper, 
$n \leq 10^4~{\rm cm}^{-3}$.

The procedure to calculate a whole set of equations is explained
in \S\ref{ssec:outline}.

%-------------------------------------------%
\subsubsection{Radiative Transfer Equations}
\label{sssec:rtrans}
%-------------------------------------------%

%---------------------------------------------------------------------------
\begin{table}[t]
%\caption{} 
\label{tb:phrate}
\begin{center}
Table 1. Properties of massive stars with solar metallicity \citep{DFS98}
\\[3mm]
\begin{tabular}{cc|ccc}
\hline
$T_{\rm eff}$ ($10^4$ K)$^a$  &    $M_*$ ($M_\odot)^b$  
                   & $\log( S_{\rm UV}({\rm s}^-1)^c )$
                   & $\log( S_{\rm FUV}({\rm s}^-1)^d )$
                   & $\log( S_{\rm FUV}/S_{\rm UV})$
\\
\hline
\hline
5.0 & 101.3 &   49.89   &  49.54  & -0.35 \\
4.0 &  40.9 &   48.78   &  48.76 &  -0.02 \\
3.3 &  19.0 &   47.75   &  48.21 & 0.54 \\
2.4 &  11.7 &   45.38   &  47.39 & 2.01 \\
\hline  
\end{tabular}

\noindent
\end{center}
a : effective temperature,  b : stellar mass,  c, d : emission rates of
UV and FUV photons    
\end{table}
%---------------------------------------------------------------------------

We solve the radiative transfer of both UV and FUV photons.
UV photons photoionize atoms in the H~II region and heat up
the gas by the photoionization.
FUV photons photodissociate molecules in the PDR and heat up
the gas mainly by the ionization of the dust grains
(photoelectric heating). 

We solve the UV ($h\nu >$13.6~eV) and FUV (11.0~eV $< h\nu <$ 13.6~eV) 
radiative transfer separately to avoid the time-consuming calculation. 
The UV radiative transfer equation is,
\begin{equation}
\label{eq:uvtr}
\frac{1}{r^2} \frac{\partial (r^2 F_{\rm \nu, UV})}{\partial r} 
= - n 
    (1 - X_{\rm H^+}) \sigma_{\rm \nu, UV} F_{\rm \nu, UV},
\end{equation}
where $F_{\rm \nu, UV}~d\nu$ is the UV photon number flux at
the frequency range, $\nu \to \nu + d\nu$
and $\sigma_{\rm \nu, UV}$ is the cross section for the photoionization
of hydrogen atoms.  
Equation (\ref{eq:uvtr}) is solved as
\begin{equation}
\label{eq:juv}
F_{\rm \nu, UV} = \frac{S_{\rm \nu, UV}}{4 \pi r^2}
                  \exp ( - \tau_{\rm \nu, UV} ) _,
\end{equation}
where $S_{\rm \nu, UV}~d\nu$ is 
the photon number luminosity from the central star at a 
frequency range of $\nu \to \nu + d\nu$, and $\tau_{\rm \nu,UV}$
is the opacity written as, 
\begin{equation}
\tau_{\rm \nu, UV} = \int 
                       n (1-X_{\rm H^+}) \sigma_{\rm \nu, UV}
                       ~dr_.
\end{equation} 
\citet{DFS98} provide the frequency-integrated photon number
luminosities of massive stars in UV and FUV range. 
They have calculated with the line-bracketed LTE atmosphere 
models by \citet{Kr93}.
We use their results for the solar metallicity listed in Table 1.
Note that some recent works present slightly larger values of $S_{\rm UV}$ 
and $S_{\rm FUV}$. For example, $S_{\rm UV}$ given by 
\citet{MSH05} is larger by a factor of 
$\sim 1.5~(3.0)$ for $20~M_\odot$ 
($40~M_\odot$) star than the adopted values. We have calculated the ratio, 
$S_{\rm UV}/S_{\rm FUV}$ with the model spectrum distributed by
\citet{LH03}, and found that $S_{\rm FUV}$ is also  
larger than the values in Table.1 only by a factor of less than 2. 
However, these differences by a factor of a few do not significantly 
affect our quantitative analysis. 
We approximate the spectrum of the Lyman-continuum, $S_{\rm \nu, UV}$,
by the Planck function of the effective temperature, $T_{\rm eff}$
and normalize $S_{\rm \nu, UV}$ so that the integrated UV luminosity
agrees with the values listed in Table 1.
We adopt the frequency-dependent cross section, 
\begin{equation}
\label{eq:sigmai}
\sigma_{\rm \nu, UV} = 6.3 \times 10^{-18} 
                        \left( \frac{\nu}{\nu_{\rm Lyl}} \right)^{-2.8}
                         ({\rm cm}^{2})_,
\end{equation}
where $\nu_{\rm Lyl}$ corresponds to the Lyman limit, 
$h \nu_{\rm Lyl}$ = 13.6 eV.
We do not take into account the absorption of ionizing photons by
dust inside the HII region because
of the uncertain size distribution of grains. We discuss the effect
of the dust in the H~II region in \S\ref{ssec:dusthii}. 
The diffuse UV radiation produced by the recombination is treated 
by the on-the-spot approximation.

The treatment of the FUV radiative transfer is a little more 
complicated. 
Hydrogen and carbon monoxide molecules are photodissociated
by the line absorption of energetic photons 
($11~{\rm eV} < h \nu < 13.6~{\rm eV}$).
Lower energetic photons ($h \nu < 11~{\rm eV}$) also contribute
to some other processes. 
For example, photons of $h \nu > 6$~eV (typical work function of grains)  
contribute to the photoelectric heating and the dust recombination
cooling \citep{BT94}.
Below, we solve the transfer of three types of FUV flux separately,
which are for the photodissociation of H$_2$ molecules, CO
molecules, and for other microprocesses. 

We use the FUV flux, $F_{\nu, {\rm H_2}}$ to calculate
the photodissociation rate of hydrogen molecules with the 
self-shielding effect. The radiative transfer equation 
for $F_{\nu, {\rm H_2}}$ is,
\begin{equation}
\label{eq:fuvtr}
\frac{1}{r^2} \frac{\partial (r^2 F_{\rm \nu, {\rm H_2}})}{\partial r}
= - n ( 0.5 X_{\rm H_2} \sigma_{\rm \nu, H_2} 
      + \sigma_{\rm d} )  F_{\rm \nu, {\rm H_2}} , 
\end{equation}
where $\sigma_{\nu, {\rm H_2}}$ and $\sigma_{\rm d}$ are 
the cross section of the photodissociation of hydrogen molecules
and the dust absorption, $X_{\rm H_2}$ is the
molecular ratio defined as, 
$X_{\rm H_2} \equiv 2 n_{\rm H_2}/n = 1-X_{\rm H^+}-X_{\rm H}$.
The dust attenuation law for the H$_2$ photodissociating photons
is $\exp (-2.5 A_{\rm V})$ \citep{TH85}, and we adopt
the corresponding cross section, 
$\sigma_{\rm d} = 1.2 \times 10^{-21}~{\rm cm}^2$. 
Equation (\ref{eq:fuvtr}) is solved as,
\begin{equation}
F_{\nu, {\rm H_2}} = \frac{S_{\nu, {\rm H_2}}}{4 \pi r^2}
                     \exp ( -  \tau_{\nu, {\rm H_2}} ) ,
\end{equation}
where $S_{\nu, {\rm H_2}}~d\nu$ is the FUV photon number luminosity
at $\nu \to \nu + d\nu$ and $\tau_{\rm H_2, \nu}$ is the optical depth
given by,
\begin{equation}
\tau_{\rm \nu, H_2} = \int n~ 
                      (  0.5 X_{\rm H_2} \sigma_{\rm \nu, H_2}
                       + \sigma_{\rm d} )
                      ~dr_.
\end{equation}
The photodissociation of H$_2$ molecules occurs via a so-called
``Solomon process'' \citep{SW67}.
We assume that all H$_2$ molecules
are at the rotational-vibrational levels of $v = 0$, $J = 0$
(ortho) or $v=0$, $J=1$ (para) of the ground electronic level. 
The ortho-to-para ratio is assumed to be a typical value, 3:1, 
which is the limit for the high temperature 
\citep{HS70}. We consider only transitions via Lyman-bands, which 
are the main channels for the ``Solomon process''. 
In the Lyman-band transitions, the rotational quantum number
changes as $\Delta J = -1$ (P transition) or 
$\Delta J = +1$ (R transition). Therefore, 
only 3 types of transitions are possible for each excited vibrational
level in $B^1\Sigma^+_u$, $v'$ ; 
$(v,J)=(0,0) \to (v',1)$, $(0,1) \to (v',0)$, and
$(0,1) \to (v',2)$. 
The vibrational quantum number, $v$ and $v'$, can
change freely. For the line profile for each transition, we adopt
a Voigt profile,
\begin{equation}
\label{eq:voigt}
\phi_j(\nu) = \frac{\gamma_j}{4 \pi^2 \Delta \nu_{\rm D}}
              {
               \int_{-\infty}^{+\infty} 
\frac{\exp( -y^2 )}{(v-y)^2 + (\gamma_j/4 \pi \Delta \nu_{\rm D})^2}~dy
               }_,
\end{equation}
where $\gamma_j$ is the natural line width, $\Delta \nu_{\rm D}$ is the 
Doppler width, and $v = (\nu - \nu_0)/\Delta \nu_{\rm D}$, where
$\nu_0$ is the frequency of the line center. 
In our calculation, we use the constant temperature, $T=100$~K 
to calculate the Voigt profile. 
We have confirmed that our numerical results are not sensitive
to $T$ or $\Delta \nu_{\rm D}$. 
As noted by \citet{HWS71}, the column density of H$_2$ 
molecules easily exceeds $10^{19}$~(cm$^{-2}$), 
then the Lorentz wings dominate the absorption of FUV photons 
rather than Doppler cores (see Fig. \ref{fig:h2rep} below).
The frequency-dependent cross section of the line is,
\begin{equation}
\label{eq:voigt_sg}
\sigma_j(\nu) = \frac{1}{\sqrt{\pi}} \frac{\sigma_j}{\Delta \nu_{\rm D}}
                \phi_j(\nu)_.
\end{equation}
The total cross section, $\sigma_j$ is given by 
$\sigma_j = \pi e^2/m_e c \cdot f_j$, where $f_j$ is the
oscillator strength. Numerical values of $f_j$, $\gamma_j$, and $\nu_0$
are taken from \citet{AD70}, \citet{Ab92}, and \citet{Ab93}.
Although there are about 20 sets of three lines of Lyman-bands in 
11.0~eV $< h \nu <$ 13.6~eV, we approximate these many lines by
adopting a representative set of lines. We pick up three 
$v' = 12$ lines. The logarithmic grids are set around these lines 
(12.82~eV $< h\nu <$ 12.93~eV) so that the resolution becomes 
higher around each line center.
We consider the constant radiation field, $S_{\rm \nu, H_2}$ in this 
frequency range and normalize this so that the integrated luminosity is
equal to the value listed in Table 1.
The absorption cross section for Lyman-bands is calculated taking
into account the line overlap,
\begin{equation}
\label{eq:sigh2}
\sigma_{\rm \nu, H_2} = (1-f_{\rm rem})\sum_j h_j \sigma_j(\nu),
\end{equation}
(see Fig. \ref{fig:h2rep})
where $h_j$ is the ratio of ortho or para H$_2$ molecules
and $f_{\rm rem}$ is the probability of the reemission.
We adopt $f_{\rm rem} = 0.14$ following \citet{Sl78}.
About another 15\% of the pumped H$_2$ molecules decay to be
dissociated (see \S\ref{ssec:chem}). 
We assume that the other pumped molecules decay to the excited $(v,J)$
levels in the ground electronic state, followed by an infrared cascade. 
Fig. \ref{fig:h2rep} shows the absorption cross section of
the hydrogen molecules, $\sigma_{\rm \nu, H_2}$ and the dust grains
for which we adopted $\sigma_{\rm d}$. 
As this figure shows, the effect of line overlap of Lyman-bands works 
at the column density, $N_{H_2} \geq 10^{20}$~cm$^{-2}$.
About half of FUV photons are absorbed by dust grains rather than
H$_2$ molecules, because $\sigma_{\rm d}$ is larger than
$\sigma_{\rm \nu, H_2}$ in the feet of the Lorentz wings.

The photodissociation of CO molecules is also caused by
the line absorption of FUV photons \citep{vDB88}. 
Since the distribution of such lines and the line overlapping 
with Lyman-Werner bands are complicated, 
we use other FUV flux, $F_{\nu, {\rm CO}}$ to calculate the 
CO photodissociation rate with some shielding effects
in the minimal frequency range.
The radiative transfer equation for $F_{\nu, {\rm CO}}$ is,
\begin{equation}
\label{eq:fuvtr_co}
\frac{1}{r^2} \frac{\partial (r^2 F_{\rm \nu, {\rm CO}})}{\partial r}
= - n ( X_{\rm CO} Z_{\rm C} \sigma_{\rm \nu, CO}
      + 0.5 X_{\rm H_2} \sigma_{\rm H_2, CO} 
      + \sigma_{\rm d, CO} )  F_{\rm \nu, {\rm CO}} , 
\end{equation}
where $X_{\rm CO}$ is the molecular fraction defined as,
$X_{\rm CO} \equiv n_{\rm CO}/(Z_{\rm C}n)$, $Z_{\rm C}$ is the
abundance of C atoms, $\sigma_{\rm \nu, CO}$, 
$\sigma_{\rm H_2, CO}$, and $\sigma_{\rm d, CO}$ are 
cross sections for the CO photodissociation, shielding by H$_2$
molecules, and the dust absorption. 
Equation (\ref{eq:fuvtr_co}) is integrated as,
\begin{equation}
\label{eq:flco}
F_{\nu, {\rm CO}} = \frac{S_{\nu, {\rm CO}}}{4 \pi r^2}
                     \exp ( -  \tau_{\nu, {\rm CO}} ) ,
\end{equation}
where $\tau_{\nu, {\rm CO}}$ is,
\begin{equation}
\tau_{\nu, {\rm CO}} = \int 
                   n~(  X_{\rm CO} Z_{\rm C} \sigma_{\rm \nu, CO}
                      + 0.5 X_{\rm H_2} \sigma_{\rm H_2, CO} 
                      + \sigma_{\rm d, CO} )
                      ~dr_.
\end{equation}
\citet{vDB88} have listed about 30 lines to photodissociate
CO molecules in the energy range of 11.0~eV $< h \nu <$ 13.6~eV. 
As for the H$_2$ molecule Lyman bands, we use the typical one line, 
$\sigma_{\nu, {\rm CO}}$ to calculate
the photodissociation rate of CO molecules.
We adopt the Voigt profile given by (\ref{eq:voigt}) and 
(\ref{eq:voigt_sg}) again, and make up the representative line
with typical parameters ; $h \nu_0 = $12.925~eV ($\lambda_0 = 960 \AA$), 
$\gamma = 1.0 \times 10^{-11}~{\rm s}^{-1}$ and $f=0.01$
respectively. 
We put the logarithmic grids at 12.88~eV $< h \nu < $ 12.96~eV
to get the resolution higher around the line center.
The radiation field, $S_{\nu, {\rm CO}}$ in equation (\ref{eq:flco})
is determined in the same manner as $S_{\nu, {\rm H_2}}$.
In the PDR, CO molecules are shielded by the dust and 
H$_2$ molecules.
\citet{Le96} have calculated the shielding function
by the dust and H$_2$ molecules for the CO photodissociation rate.
We have found that their results (Table 11) can be approximated
with some analytic functions.
Their dust attenuation law can be fitted with
$\exp(-3.5 A_{\rm V})$ within a factor of 2 at $A_{\rm V} < 7$.
Similarly, their H$_2$ molecule shielding function can be fitted with 
$\exp(-N_{\rm H_2}/1.6 \times 10^{21}~{\rm cm}^{-2})$ 
within 20\% error at $N_{\rm H_2} < 8 \times 10^{21}~{\rm cm}^{-2}$. 
Therefore, we adopt the corresponding cross sections in our
calculations, $\sigma_{\rm d, CO} = 1.7 \times 10^{-21}~{\rm cm}^2$
and $\sigma_{\rm H_2, CO} = 6.25 \times 10^{-22}~{\rm cm}^2$ respectively. 
Fig.\ref{fig:corep} shows the cross sections for the CO 
photodissociation rate. As this figure shows, the self-shielding 
effect is not so significant owing to the small abundance of C atoms.
The dust cross section is largest in most of the energy range,
which means that CO molecules are mainly shielded by the dust absorption.

Finally, we define the frequency-integrated FUV photon number flux
normalized with the Habing's unit, $G_{\rm FUV}$ \citep{Hb68}.
We assume that $G_{\rm FUV}$ is attenuated only by the dust absorption,
\begin{equation}
\label{eq:rtr_uv}
\frac{1}{r^2} \frac{\partial (r^2 G_{\rm FUV})}{\partial r} 
= - n \sigma_{\rm d} G_{\rm FUV} ,
\end{equation}
which is solved as,
\begin{equation}
\label{eq:fuvG}
G_{\rm FUV} = \frac{1}{F_{\rm H}}  
              \frac{S_{\rm \nu, FUV}}{4 \pi r^2}
                  \exp ( - \tau_{\rm d} ) _,
\end{equation}
where $F_{\rm H}$ is the Habing field, 
$F_{\rm H}= 1.21 \times 10^7~{\rm cm}^{-2} {\rm s}^{-1}$ 
\citep{Hb68, BD96}, and $\tau_{\rm d}$ is the dust optical depth 
defined as,
\begin{equation}
\tau_{\rm d} = \int
               n \sigma_{\rm d}~dr_.
\end{equation} 
We adopt $\sigma_{\rm d} = 1.2 \times 10^{-21}~{\rm cm}^{-2}$,
which is the average value between 1000 and 2000~$\AA$ and
corresponds to the attenuation law of $\exp(-2.5 A_{\rm V})$
\citep{Rb91}.
As mentioned above, we assume $\sigma_{\rm d} = 0$ in the H~II region
for simplicity. We use this flux, $G_{\rm FUV}$ to calculate the photoelectric 
heating and dust recombination cooling rate \citep{BT94}, 
reformation rate of CO molecules \citep{NL97}, 
and dust temperature \citep{HTT91}.

%------------------------------------------%
\subsubsection{Chemical Reaction Equations}
\label{ssec:chem}
%------------------------------------------%

%---------------------------------------------------------------------------
\begin{table}[tbp]
%\caption{Included Chemical Processes}
\label{tb:ch}
\begin{center}
Table 2. Included Chemical Processes \\[3mm]
%{\scriptsize
\begin{tabular}{l|l|l|l}
\hline
species & reactions  & reference & note \\
\hline
\hline
               & photoionization         & see \S\ref{ssec:chem} &  \\
 H $\to$ H$^+$ & collisional ionization  & 1        &  \\
               & cosmic-ray              & 2        &  \\
\hline
 H$^+$ $\to$ H & recombination & 3    &  \\
\hline
 H$_2$ $\to$ H$^+$  & photoionization & see \S\ref{ssec:chem} &  \\
                    & cosmic-ray      & 2          &  \\
\hline
 H$_2$ $\to$ H  & photodissociation       & see \S\ref{ssec:chem} &  \\
                & cosmic-ray              & 2          &  \\
\hline
 H $\to$ H$_2$  & reformation        & 4  &             \\
                & gas-phase reaction & 4  &  $\dagger$1 \\              
\hline
 CO $\to$ C$^+$  & photodissociation & see \S\ref{ssec:chem} &  \\
\hline
 C$^+$ $\to$ CO  & reformation       & 5  &  \\
\hline
\end{tabular}
%}
\noindent
\end{center}
REFERENCES---(1) Tenorio-Tagle 1986, (2) Wolfire et al. 1995,
(3)     , (4) Hollenbach \& McKee 1979 (HM79),
(5) Nelson \& Langer 1997
\ NOTES---$\dagger$1 : 
we use the rate by HM79 for reactions; 
H + e $\rightarrow$ H$^-$ + $\gamma$ and 
H$^-$ + H $\rightarrow$ H$_2$ + e \
\end{table}
%---------------------------------------------------------------------------

Non-equilibrium reaction equations are solved for the 
species of e, H, H$^+$, H$_2$, C$^+$, and CO. 
The ionization rate of O is assumed to be the same as that of
H. We adopt the solar elemental abundances with some depletions, 
$Y_{\rm He}= 0.1$, $Z_{\rm O} = 5.4 \times 10^{-4}$, and
$Z_{\rm C} = 2.3 \times 10^{-4}$ \citep{HM89}.

We solve the minimal set of chemical reactions listed in
Table 2.
The main reactions in the H~II region are the photoionization
by the UV photons from the central star and the recombination.
In the PDR, the FUV photons photodissociate H$_2$ and CO molecules. 
The reformation of H$_2$ molecules mainly occurs on the surface 
of dust grains. 
The dissociation and reformation of CO molecules is approximated with the
direct process between CO and C$^+$ \citep{NL97}.

Below, we show the photoionization and photodissociation rates,
which relate to the transfer equations explained in \S\ref{sssec:rtrans}. 
The photoionization rates of H and H$_2$, 
photodissociation rates of H$_2$ and CO molecules are,
\begin{equation}
\frac{dX_{\rm H^+}}{dt} = - \frac{dX_{\rm H}}{dt} = 
                 X_{\rm H}
                 \int_{\nu_{\rm Lyl}}^\infty
                 \sigma_{\rm \nu, UV}~F_{\rm \nu, UV}
                 ~d\nu' \ ,
\end{equation}
\begin{equation}
\label{eq:phih2}
\frac{dX_{\rm H^+}}{dt} = - 2 \frac{dX_{\rm H_2}}{dt} 
               = 2 X_{\rm H_2}
                 \int_{\nu_{\rm Lyl}}^\infty
                 \sigma_{\rm \nu, UV}~F_{\rm \nu, UV}
                 ~d\nu' \ ,
\end{equation}
\begin{equation}
\label{eq:phdh2}
\frac{dX_{\rm H}}{dt} = - 2 \frac{dX_{\rm H_2}}{dt} =
                    2 X_{\rm H_2} 
                    \int_{\nu_-}^{\nu_+}
                    \frac{p_{\rm dis}}{1 - f_{\rm rem}} 
                    \sigma_{\rm \nu, H_2}~F_{\rm \nu, H_2}
                    ~d\nu' \ ,
\end{equation}
\begin{equation}
\label{eq:phdco}
\frac{dX_{\rm C^+}}{dt} = - \frac{dX_{\rm CO}}{dt} 
                 = X_{\rm CO} 
                  \int_{\nu_-}^{\nu_+} 
                  \sigma_{\rm \nu,CO}~F_{\rm \nu, CO}
                  ~d\nu' \ ,
\end{equation}
where $p_{\rm dis}$ in equation (\ref{eq:phdh2}) is the probability
that the pumped H$_2$ molecules are photodissociated, decaying to
$X^1\Sigma_g^+$. We adopt the average value, $p_{\rm dis} = 0.15$
\citep{DB96}. The integration in equations (\ref{eq:phdh2}) and
(\ref{eq:phdco}) are done in the considered frequency range
around our representative lines. 
As equation (\ref{eq:phih2}) shows, we assume that
if UV photons ($> 13.6$~eV) reach the molecular region,
one H$_2$ molecule is directly ionized to two H$^+$ ions 
as two H atoms are photoionized.

%%%%%%%%%%%%%%%%%%%%%%%%%%%%%%%
\subsection{Thermal Processes}
\label{ssec:thm}
%%%%%%%%%%%%%%%%%%%%%%%%%%%%%%%

%---------------------------------------------------------------------------
\begin{table}[tbp]
%\caption{Included Thermal Processes}
\label{tb:th}
\begin{center}
Table 3. Included Thermal Processes \\[3mm]
%{\scriptsize
\begin{tabular}{l|c|l|l|l}
\hline
 & region & process & reference & note \\
\hline
\hline
        & HII & H photoionization       & 1 etc. &  \\
\cline{2-5}
        &     & Photoelectron from dust & 2 & \\
Heating & PDR & H$_2$ photodissociation & 3 & \\
        &     & H$_2$ reformation       & 3 & \\
        &     & Cosmic-ray              & 4 & \\
\hline
        &     & H recombination        & 1 etc. &  \\
        &     & Lyman-$\alpha$         & 1 & \\
        & HII & OI  (63.0$\mu$m)       & 5 & \\
        &     & OII (37.29$\mu$m)      & 5 & \\
        &     & CII (23.26$\mu$m)      & 5 & \\
Cooling &     & Collisional ionization & 6 &  \\
\cline{2-5}
        &     & OI  (63.1 $\mu$m)      & 5, 7 & $\dagger$1 \\
        &     & CII (157.7$\mu$m)      & 5, 7 & $\dagger$1 \\
        & PDR & H$_2$ rot/vib excitation & 3, 8 &  \\
        &     & CO rot/vib excitation    & 3    &  \\
        &     & Dust recombination       & 2    &  \\
        &     & Collisional dust-gas heat transfer   & 5  & $\dagger$2 \\
\hline 
\end{tabular}
%}
\noindent
\end{center}
REFERENCES---(1) Spitzer 1978, (2) Bakes \& Tielens 1994,
(3) Hollenbach \& McKee 1977, (4) Shull \& Van Steenberg 1985,
(5) Hollenbach \& McKee 1989, (6) Tenorio-Tagle 1986, 
(7) Tielens \& Hollenbach 1985, 
(8) Galli \& Palla 1998. \ NOTES---$\dagger$1 : 
these can be a heating process when the background radiation is 
greater than the source function. \
$\dagger$2 : this can be a heating process when the 
dust temperature is higher than the gas temperature. 
\end{table}
%---------------------------------------------------------------------------

We include 5 heating processes and 12 cooling processes in the 
energy equation. These processes are listed in Table 3.
In Table 3, heating/cooling processes are divided into
processes dominant in the H~II region or PDR. In the H~II region, 
the included heating process is photoionization heating of hydrogen atoms. 
The main cooling processes in the H~II region are recombination cooling 
and some meta-stable atomic lines (e.g., [O~II] 3729 \AA, [C~II] 2326 \AA).
We do not include the photoelectric heating in the H~II region,
because this is sensitive to the uncertain abundance of small grains.
We do not include the line cooling with highly ionized ions
(e.g., [O~III], [C~III] etc.) either. We are interested in the boundary
region between the H~II region and the molecular cloud, and thus we
do not solve the detailed ionization structure in the H~II region.  

In the PDR, the main heating process is photoelectric heating.
We separately list the dust recombination cooling, which exceeds the
photoelectric heating at high temperatures 
\citep[$T > 10^3$~K,][]{BT94}.
In our calculation, cosmic-ray heating is important only in the outer
molecular region, where the FUV flux from the central star
is significantly reduced by geometrical dilution and dust
extinction. The main cooling processes in the PDR
are fine-structure atomic lines of [O~I] 63.1$\mu$m and [C~II] 157.7$\mu$m,
which are the characteristic lines of the PDR \citep[e.g.,][]{HT99}. 
The [O~I] 63.1$\mu$m is more efficient than [C~II] 157.7$\mu$m
at high density ($n \geq 10^4~{\rm cm}^{-3}$), and it is the
reverse at the low density ($n \leq 10^2~{\rm cm}^{-3}$).   
We do not include [CI] 370$\mu$m and 609$\mu$m transitions
because we adopt the approximate treatment that CO molecules are
directly photodissociated to C$^+$ \citep{NL97}. 
The rotational transitions of CO molecules are the primary cooling 
process in the molecular region. 
Collisional heat transfer between gas and dust is important
when the density is very high ($> 10^5$ cm$^{-3}$). 
If dust temperature is higher than the gas
temperature, this process heats up the gas.

%%%%%%%%%%%%%%%%%%%%%%%%%%%%%%
\subsection{Dust Temperature}
\label{subsec:tdust} 
%%%%%%%%%%%%%%%%%%%%%%%%%%%%%%

We calculate the dust temperature following the approximation method 
by \citet{HTT91}. In our calculation, several thermal processes
depend on the dust temperature.
Dust IR emission affects the level population of atoms and cooling 
rates with line transitions (e.g., [OI] 63.1$\mu$m and [C~II] 157.7$\mu$m). 
If the intensity of the IR dust emission
at the line frequency exceeds the source function of the line, this
transition works as a heating process rather than a cooling process.  
The collisional dust-gas heat transfer rate is
proportional to the difference between the gas and
dust temperature.

%%%%%%%%%%%%%%%%%%%%%%%%%%%%%%%%%%%%%%%%
%%%%%%%%%%%%%%%%%%%%%%%%%%%%%%%%%%%%%%%%
\section{Calculation Procedures}
%%%%%%%%%%%%%%%%%%%%%%%%%%%%%%%%%%%%%%%%
%%%%%%%%%%%%%%%%%%%%%%%%%%%%%%%%%%%%%%%%

%%%%%%%%%%%%%%%%%%%%%%
\subsection{Outline}
\label{ssec:outline}
%%%%%%%%%%%%%%%%%%%%%%

Our numerical scheme for the hydrodynamics is based 
on the 2nd-order Lagrangian Godunov method 
\citep[see, e.g.,][]{vL97}.
First, we explicitly integrate the continuity equation and the momentum 
equation to obtain the new position of the cell, $r_j^{~n+1}$, gas density, 
$\rho_j^{~n+1}$, and velocity, $u_j^{~n+1}$, where 
the indices $j$ and $n$ refer to different steps in space and time
respectively. Next, we iteratively solve the chemical rate equations 
and the energy equation. The energy equation
(\ref{eq:egyeq}) is transformed with equations (\ref{eq:mmtm}), 
(\ref{eq:eos}) and (\ref{eq:rhon}) as, 
\begin{eqnarray}
\label{eq:egy_ite}
\frac{d T}{d t} &+& 
\frac{T}{2 w}
\left(    \frac{d X_{\rm H}}{d t} 
      + 3 \frac{d X_{\rm H^+}}{dt} \right) \nonumber \\
&+& \frac{\mu m_{\rm H} (\gamma - 1)}{k w}
\left\{  u r^2 \frac{\partial p}{\partial m} 
      -  \frac{\partial (r^2 u p)}{\partial m} 
      -  \frac{\partial (r^2 q_{\rm cond})}{\partial m}  \right\}
                                              \nonumber \\
&=& \frac{\gamma - 1}{k w} (\Gamma - \Lambda),
\end{eqnarray}
where $w$ is defined as,
\begin{equation}
w \equiv 0.5(1 + X_{\rm H} + 3 X_{\rm H^+}) + Y_{\rm He} \ .
\end{equation}
The terms with $m$-derivative are explicitly calculated  
before the iteration.
The different rate equations and energy equation include the new 
gas temperature, $T_j^{~n+1}$, dust temperature, 
${T_{\rm d,j}}^{n+1}$, and chemical compositions, ${X_{s,j}}^{n+1}$ 
($s$ : species). We determine these quantities by means of the 
Newton-Raphson method.
After the iteration converges, the gas pressure and UV/FUV flux 
are calculated using the newly obtained gas density, temperature, 
and compositions. The integration proceeds from the center to 
the outer cells.

%%%%%%%%%%%%%%%%%%%%%%%%%%%%%%
\subsection{Initial Condition} 
\label{sec:initia}
%%%%%%%%%%%%%%%%%%%%%%%%%%%%%%

The hydrodynamical expansion of the H~II region occurs
after the IF passes the initial Str\"omgren radius
(expansion phase).  We start our calculation 
when the H~II region is still in the previous formation phase.
In the formation phase, the timescale of the system is very
short and the expanding H~II region hardly affects the gas dynamics.
At $t=0$, we set the constant mass density and no initial velocity
field. We set the initial H~II region 
($X_{\rm H^+}= X_{\rm C^+}=1$, $T=10^4$~K) at $r < R_{\rm st}/5$.
The outer initial PDR ($X_{\rm H}= X_{\rm C^+}=1$, $T=100$~K) and the 
molecular region ($X_{\rm H_2}= X_{\rm CO}=1$, $T=10$~K)
is set at $R_{\rm st}/5 < r < 2 R_{\rm st}/5$ and $r > 2 R_{\rm st}/5$
respectively.

%%%%%%%%%%%%%%%%%%%%%%%%%%%%%%%%%%%%%%%%%%%%%%%%%%%%%%%%%%%
\subsection{Time Step} 
\label{sec:tstep}
%%%%%%%%%%%%%%%%%%%%%%%%%%%%%%%%%%%%%%%%%%%%%%%%%%%%%%%%%%%

The time step of the calculation, $\Delta t$, is constrained by
some conditions.
The first one is the normal Courant condition,
\begin{equation}
\Delta t_{{\rm c},j} = \frac{\Delta r_j}{C_{{\rm s},j}} \ ,
\end{equation}
where $C_{{\rm s},j}$ is the sound speed at the j-th cell. 
We include the thermal conduction explicitly
in the energy equation, and we have another Courant condition,
\begin{equation}
\Delta t_{{\rm cond},j} = \frac12 \frac{p_j}{K_j T_j (\gamma-1)}
                        (\Delta r_j)^2 ,
\end{equation}
where $K_j$ is the conductivity calculated with equation
(\ref{eq:cduc}) to satisfy the Field condition.
Besides these Courant conditions, $\Delta t$ is constrained by
the velocity of the IF.
In the early formation phase of the H~II region, the IF expands
supersonically. Since the ionization rate and the temperature
change sharply at the IF, $\Delta t$ must be short enough
that the IF advances less than the width of one cell, $\Delta r_j$
in one time step, $\Delta t$. 

We choose the minimum time step from among the above time steps to
determine $\Delta t$. In the early formation phase, $\Delta t$ is
mainly constrained by the expanding velocity of the IF. In the
next expansion phase, when the SF emerges, the Courant conditions
limit the time step.

%---------------------------------------------------------------------------
\begin{table}[t]
%\caption{Model Parameters}
\label{tb:md}
\begin{center}
Table 4. Model Parameters \\[3mm]
%{\scriptsize
\begin{tabular}{l|cccc}
\hline
Model  & $M_*~(M_\odot)^a$ & 
         $n_{\rm H,0} ({\rm cm^{-3}})^b$ &
         $R_{\rm st}~({\rm pc})^c$ & $t_{\rm dyn}~({\rm Myr})^d$ \\
\hline
\hline
S101      & 101.3 & $10^3$   & 1.27  &  0.12   \\
S41$^e$   & 40.9  & $10^3$   & 0.56  &  0.05   \\
LD-S41    & 40.9  & $10^2$   & 2.53  &  0.23   \\
HD-S41    & 40.9  & $10^4$   & 0.12  &  0.011  \\
S19       & 19.0  & $10^3$   & 0.25  &  0.023  \\
S12       & 11.7  & $10^3$   & 0.04  &  0.004  \\
\hline                           
\end{tabular}
%}
\noindent
\end{center} 
a : mass of the central star, b : number density of the ambient gas,
c : initial Str\"omgren radius, d : dynamical time, 
$R_{\rm st}/C_{\rm HII}$, where $C_{\rm CII}$ is the sound speed at 
$T = 10^4$~K. e : Our model for Sh104 (Paper I).
\end{table}
%---------------------------------------------------------------------------

%%%%%%%%%%%%%%%%%%%%%%%%%%%%%%%%%%%%%%%%%%%%%%%%%%%%%%%%
\section{Dependence on Ambient Number Density}
\label{sec:ndep}
%%%%%%%%%%%%%%%%%%%%%%%%%%%%%%%%%%%%%%%%%%%%%%%%%%%%%%%%%

%--------------------------------%
\subsection{Models}
\label{ssec:mdl_n}
%--------------------------------%

In this section, we show how the time evolution of the H~II region,
PDR, and the swept-up shell changes with different ambient
number densities.
For comparison, we calculate the expansion of the H~II region
around the $41 M_\odot$ star, which is the central star of Sh104
modeled in Paper I, but with different ambient number densities.
As we have adopted the ambient number density of
$n_{\rm H,0} = 10^3~{\rm cm}^{-3}$ in Paper I, we show the results
of the higher density case,  $n_{\rm H,0} = 10^4~{\rm cm}^{-3}$
and the lower density case,  $n_{\rm H,0} = 10^2~{\rm cm}^{-3}$
below. The initial Str\"omgren radius and the dynamical time for
the calculated models are listed in Table 4.
All the other conditions are the same as those in our model of Sh104 
(S41 in Table 4) except for the number density. 

%---------------------------------%
\subsection{Scaling Relations}
\label{ssec:scl_n}
%---------------------------------%

First, we show some scaling relations that are very useful to
interpret the numerical results.  
The initial Str\"omgren radius depends on the ambient number density, 
and other relevant quantities are also scaled as,
\begin{equation}
\label{eq:rst_nsc}
R_{\rm st} \propto n^{-2/3} ,
\end{equation}
\begin{equation}
\label{eq:tdy_nsc}
t_{\rm dyn} = R_{\rm st}/C_{\rm HII} \propto n^{-2/3} ,
\end{equation}
\begin{equation}
\label{eq:msh_nsc}
M_{\rm sh} \propto n R_{\rm st}^3 \propto n^{-1} ,
\end{equation}
\begin{equation}
\label{eq:sig_nsc}
N_{\rm sh} \propto n R_{\rm st} \propto n^{1/3} ,
\end{equation}
\begin{equation}
\label{eq:jfuvi_nsc}
F_{\rm FUV,i} \propto S_{\rm FUV}/R_{\rm st}^2 \propto n^{4/3} ,
\end{equation}
where $M_{\rm sh}$ and $N_{\rm sh}$ is the mass
and the column density of the shell, $F_{\rm FUV,i}$ 
is the incident FUV flux at the IF at a given $t/t_{\rm dyn}$.
To derive the above relations, we assume that most of the 
swept-up gas remains in the shell, which is a good approximation
for expansion in a homogeneous ambient medium.
As (\ref{eq:rst_nsc}) and (\ref{eq:tdy_nsc}) show, 
the larger H~II region expands in the
lower density ambient medium over a longer timescale.
Equation (\ref{eq:msh_nsc}) shows that
the swept-up mass is larger with the lower density medium.
The shell structure is mainly determined by
the column density of the shell, $N_{\rm sh}$ 
and the FUV flux at the IF, $F_{\rm FUV,i}$.
The detailed properties of the shell are explained with
equation (\ref{eq:sig_nsc}) and (\ref{eq:jfuvi_nsc}) 
in \S\ref{ssec:rlt_n}.

%---------------------------------------------------%
\subsection{Results of Numerical Calculations}
\label{ssec:rlt_n}
%----------------------------------------------------%

\begin{figure}[tb]
\begin{center}
\includegraphics[width=0.6\hsize]{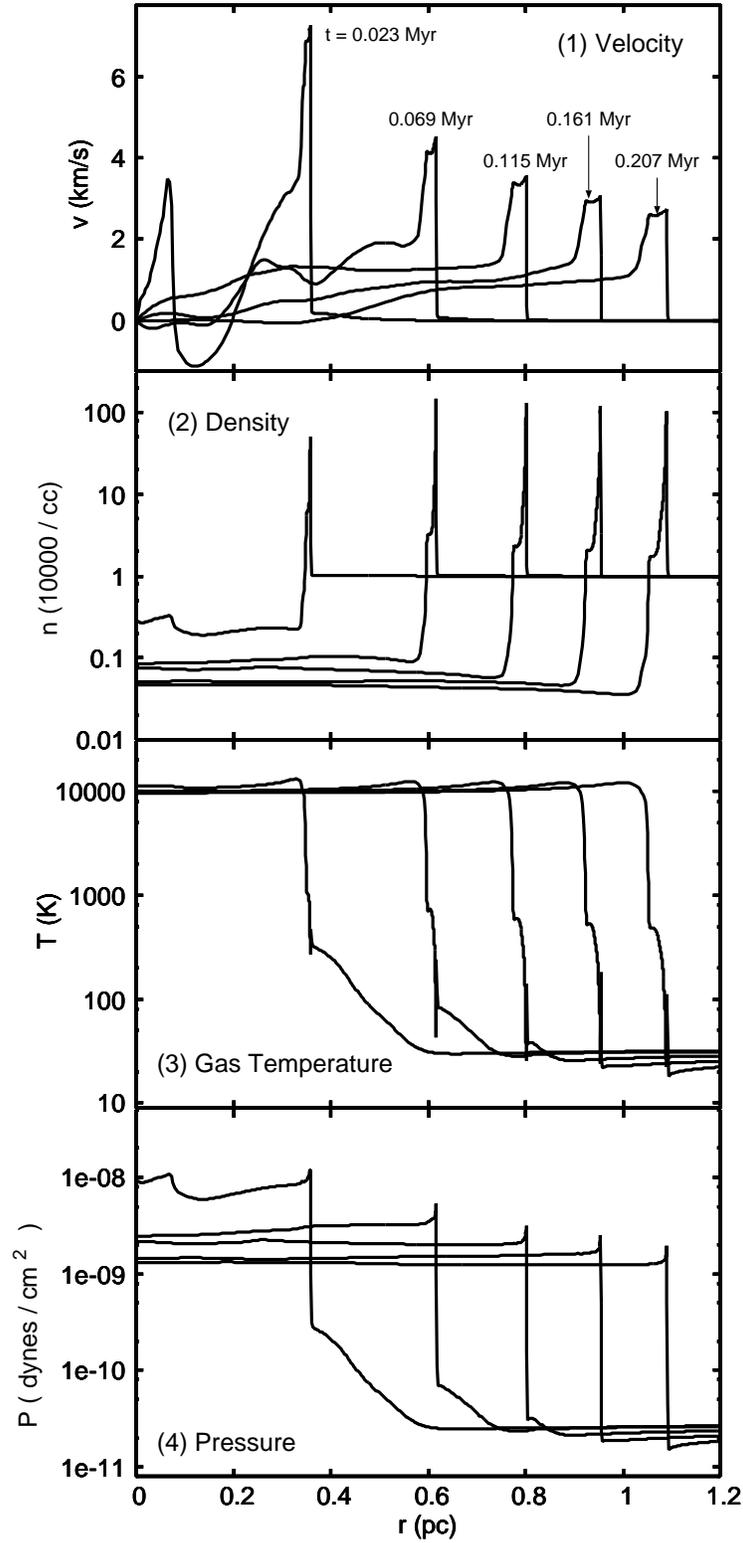}
\caption{The snapshots of the gas-dynamical evolution of model
HD-S41. In each panel, five snapshots
represent the profiles at $t = 0.023$, 0.069, 0.115, 0.161 and
0.207~Myr respectively.
 }
\label{fig:hev_5000}
\end{center}
\end{figure}

\begin{figure}[tb]
\begin{center}
\includegraphics[width=0.6\hsize]{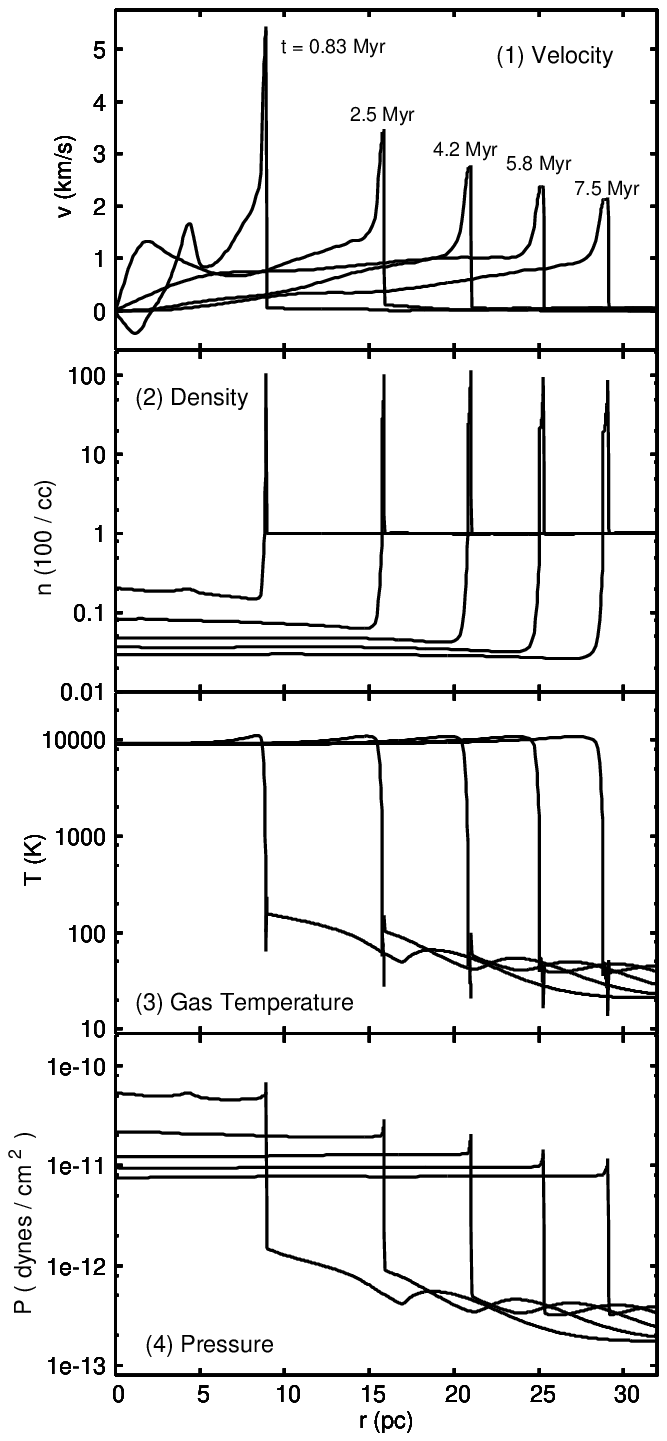}
\caption{Same as Fig\ref{fig:hev_5000} but for model LD-S41. 
In each panel, five snapshots represent the profiles at 
$t = 0.83$, 2.5, 4.2, 5.8 and 7.5~Myr respectively.}
\label{fig:hev_50}
\end{center}
\end{figure}

Figures \ref{fig:hev_5000} and \ref{fig:hev_50} show the hydrodynamical
evolution of models HD-S41 and LD-S41 respectively.
Since the Str\"omgren radius is smaller for the higher ambient density,
the H~II region expands to the region of $\sim 1$~pc in HD-S41,
and $\sim 20$~pc in model LD-S41 in $t \sim 20~t_{\rm dyn}$. 
As expected with equation (\ref{eq:tdy_nsc}),
the timescale of the evolution is longer for the lower density model.
The time of $20~t_{\rm dyn}$ corresponds to
$\sim 0.25$~Myr for HD-S41, and $\sim 5$~Myr for model LD-S41. 

\begin{figure}[tb]
\begin{center}
\includegraphics[width=0.8\hsize]{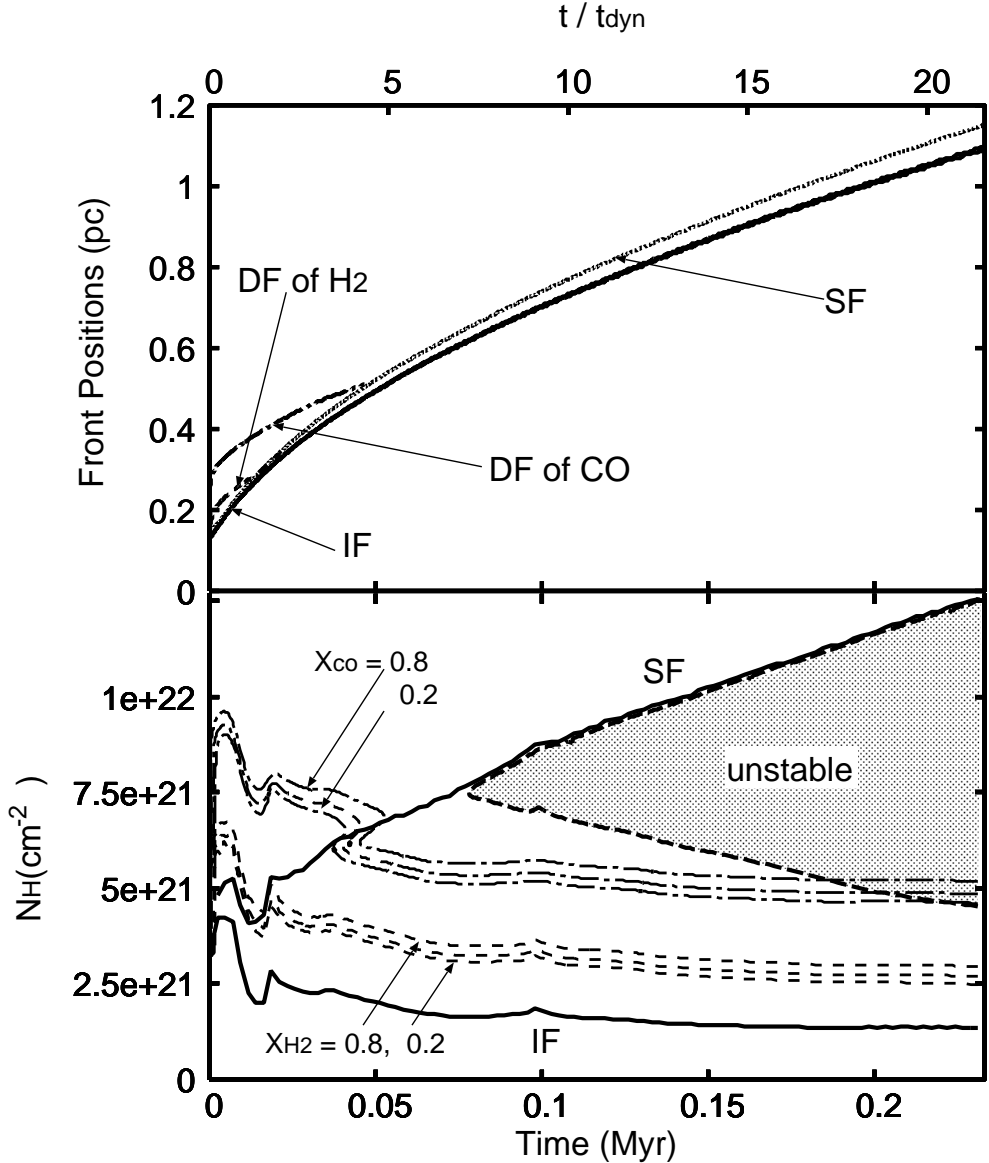}
\caption{
{\it Upper panel:} 
The time evolution of the various front positions in model HD-S41.
The solid and dotted line mean the position of the IF and SF. 
The broken (dot-solid)
line represents the DF of H$_2$ (CO), where 
$X_{\rm H_2} \equiv 2 n_{\rm H_2}/n_{\rm H_{nuc}} = 0.5$ and 
$X_{\rm CO} \equiv n_{\rm CO}/n_{\rm C_{nuc}} = 0.5$. 
We plot the position of DFs only before the SF catches up with 
each DFs. After the DFs are taken in the shell, the positions 
of DFs are close to that of the SF.
{\it Lower panel:} The time evolution of the column density of each
 region in model HD-S41, where the vertical axis, $N_{\rm H}$ is
the column density of the hydrogen nuclei included in all
 species of H$^+$, H and H$_2$ from the central star. 
 Thin contour lines represent the position where 
 $X_{\rm H_2}$ ({\it dashed contours}) and $X_{\rm CO}$ 
({\it dot-dashed contours}) = 0.2, 0.5, and 0.8. The shaded region
corresponds to the expected unstable region, where $t > (G \rho)^{-1/2}$.
}
\label{fig:fev_5000}
\end{center}
\end{figure}

\begin{figure}[tb]
\begin{center}
\includegraphics[width=0.8\hsize]{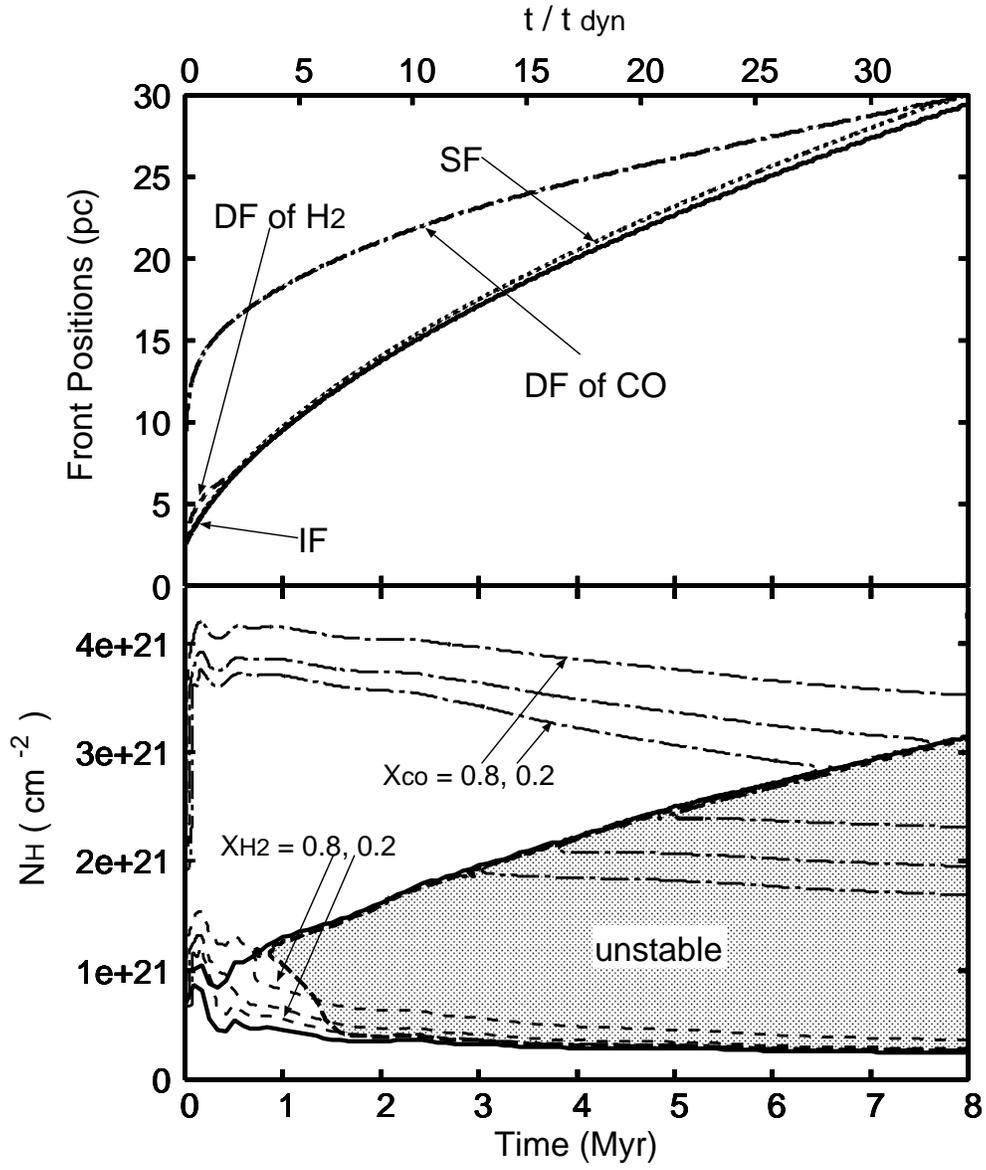}
\caption{
Same as Fig.\ref{fig:fev_5000} but for model LD-S41.
}
\label{fig:fev_50}
\end{center}
\end{figure}

The basic time evolution is the same as that of model S41 (Paper I). 
The SF emerges when the IF reaches the initial 
Str\"omgren radius. The gas density within the shell is about 
10-100 times as high as that of the ambient medium, though the 
detailed structure of the shell differs between HD-S41 and
LD-S41 (explained below). 
The time evolution after 
the SF emerges is also similar to that of model S41. 
The time evolution of the radial expansion is well approximated
with, 
\begin{equation}
\label{eq:exlaw}
R_{\rm IF}(t) = R_{\rm st}
                \left(
            1 + \frac74 \sqrt{\frac43} \frac{C_{\rm II}t}{R_{\rm st}}
                \right)^{4/7} ,
\end{equation}  
which is derived from the equation of motion of the shell,
\begin{equation}
\frac{d}{dt} \left( \frac{4 \pi}{3} R_{\rm IF}^3 \rho_0
                    \dot{R_{\rm IF}}  \right)
= 4 \pi R_{\rm IF}^2 p_{\rm II}
= 4 \pi R_{\rm IF}^2 \rho_0 C_{\rm II}^2
  \left( \frac{R_{\rm st}}{R_{\rm IF}} \right)^{3/2} ,
\label{eq:eomshell}
\end{equation}
where $\rho_0$ is the mass density of the ambient medium, and
$p_{\rm II}$ is the gas pressure in the H~II region.
Equation (\ref{eq:exlaw}) is different from the well-known 
expansion law \citep[e.g.,][]{Sp78} by a factor of $\sqrt{4/3}$,
but shows better agreement with the numerical results.  
As equation (\ref{eq:eomshell}) shows, equation (\ref{eq:exlaw}) 
does not involve the detailed structure and the chemical composition 
of the shell, which we analyze with our numerical calculations.
The swept-up mass in the shell is larger for the lower
ambient number density, as equation (\ref{eq:msh_nsc}) shows.
For example, $M_{\rm sh} \sim 10^5 M_\odot$ in model LD-S41
and $M_{\rm sh} \sim 10^3 M_\odot$ in model HD-S41 at 
$t \sim 20~t_{\rm dyn}$ respectively.
The upper panels of Fig.\ref{fig:fev_5000} and \ref{fig:fev_50}
show the time evolution of various front positions.
Initially, the DFs expand more rapidly than the SF and the IF, 
then the PDR appears in front of the SF. 
As the H~II region expands, the IF and the SF gradually
overtake the preceding DFs of H$_2$ and CO in turn. Finally the DFs
are taken up into the shell. The SF enters the molecular region,
and the molecular gas is accumulated within the shell.
The rapid reformation within the shell also helps to accelerate
the accumulation of the molecular gas.

\begin{figure}[p]
\begin{center}
\includegraphics[width=0.8\hsize]{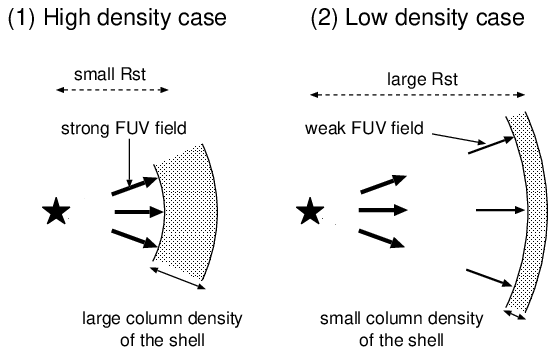}
\caption{Schematic figures for the H~II region and the swept-up
shell in models with different ambient number densities.
These figures show the snapshot at one $t/t_{\rm dyn}$
in both models.
 }
\label{fig:dscale_schem}
\end{center}
\end{figure}

The main difference among these models is the time (or $t/t_{\rm dyn}$)
when the DFs are taken into the shell.  
In the higher density model, HD-S41, the IF and the
SF catch up with both DFs of H$_2$ and CO 
by the time of $5~t_{\rm dyn}$.
In the lower density model, LD-S41, the DF of H$_2$ is quickly taken into
the shell, but the accumulation
of CO molecules is delayed. The DF of CO is taken into the shell
at $t \sim 20-30~t_{\rm dyn}$ in model LD-S41.
About 70\% (90\%) of carbon (hydrogen) atoms within the shell exist 
as CO (H$_2$) molecules in model HD-S41 at $t \sim 20~t_{\rm dyn}$, 
but 20\% (90\%) in model LD-S41 at the same time step.  
We explain the reasons for this difference below.
With the higher ambient number density, the incident
FUV flux at the IF is strong, as equation
(\ref{eq:jfuvi_nsc}) shows.  Although the FUV photons are available
to photodissociate molecules, our calculation shows that the
FUV photons are quickly shielded in model HD-S41. 
This is because the column density of the shell,
$N_{\rm sh}$ is larger for the higher density model, as
equation (\ref{eq:sig_nsc}) shows.
Fig.\ref{fig:dscale_schem} is a schematic picture showing
the density-dependence of $N_{\rm sh}$ and $F_{\rm FUV,i}$.
The incident FUV flux is attenuated enough by the dust extinction
through the shell, and the molecular gas 
is reliably protected in the outer part of the shell. 
Though the density-dependence of $N_{\rm sh}$, 
$N_{\rm sh} \propto n^{1/3}$, is weaker than that of 
$F_{\rm FUV,i}$, $F_{\rm FUV,i} \propto n^{4/3}$, the
dust absorption becomes significant with a high number density owing
to the attenuation law of $\exp(- \sigma_{\rm d} N_{\rm sh})$. 
The lower panels of Fig.\ref{fig:fev_5000} 
shows the time evolution of the column density of each region
in model HD-S41. The column density of the shell at
$t \sim 5~t_{\rm dyn}$ is already 
$N_{\rm sh} \sim 4 \times 10^{21}~{\rm cm}^{-2}$ 
($A_{\rm V, sh} \sim 2$) in this model.
In model LD-S41, on the other hand, the incident FUV flux is weak, 
but the column density of the shell is low.
Since the CO molecules are mainly shielded by the dust absorption,
it is not until $N_{\rm sh}$ becomes somewhat large
($A_{\rm V, sh} \sim 1$) that CO molecules are accumulated
in the shell.  However, even in the early phase when $A_{\rm V,sh} < 1$,
H$_2$ molecules can be quickly accumulated in the shell, 
as Fig.\ref{fig:fev_50} shows. 
This is due to the efficient self-shielding of H$_2$ molecules.
\citet{HT99} have briefly shown that H$_2$ molecules can be 
self-shielded, if $G_{\rm FUV}/n \leq 4 \times 10^{-2}$ at the IF.
In our calculations, $G_{\rm FUV}/n \sim 1.6 \times 10^{-2}$ in model
LD-S41 and $G_{\rm FUV}/n \sim 0.5$ in model HD-S41 at $t \sim 5~t_{\rm dyn}$.
Therefore, H$_2$ molecules are mainly self-shielded in
model LD-S41, and shielded by the dust in model HD-S41 respectively.
The self-shielding effect for CO molecules is not so efficient,
owing to the small abundance of carbon atoms.

\begin{figure}[p]
\begin{center}
\includegraphics[width=0.48\hsize]{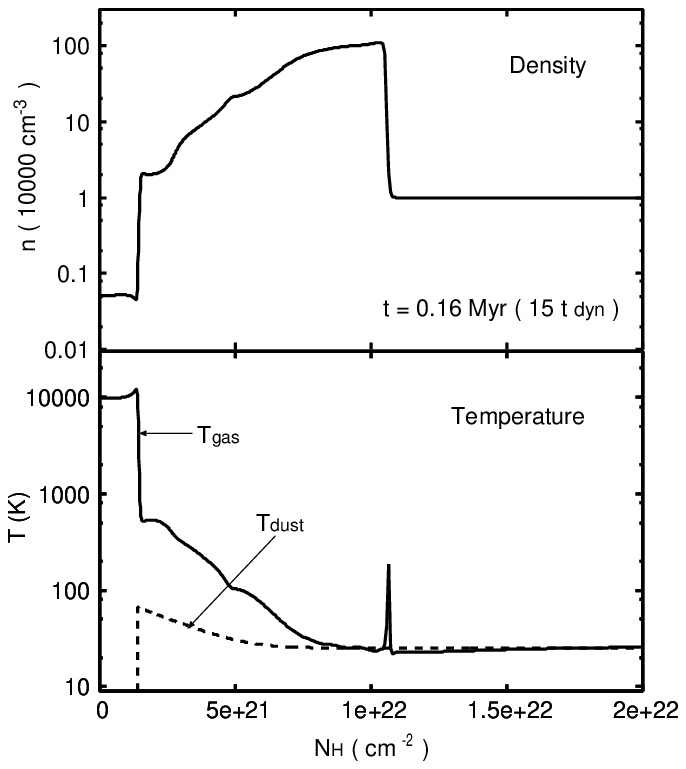}
\caption{
The density and the gas/dust temperature profiles in model HD-S41.
These snapshots are at $t = 0.161$~Myr, which corresponds to 
$\sim 15~t_{\rm dyn}$.}
\label{fig:over_5000}
\vspace{5mm}
\includegraphics[width=0.48\hsize]{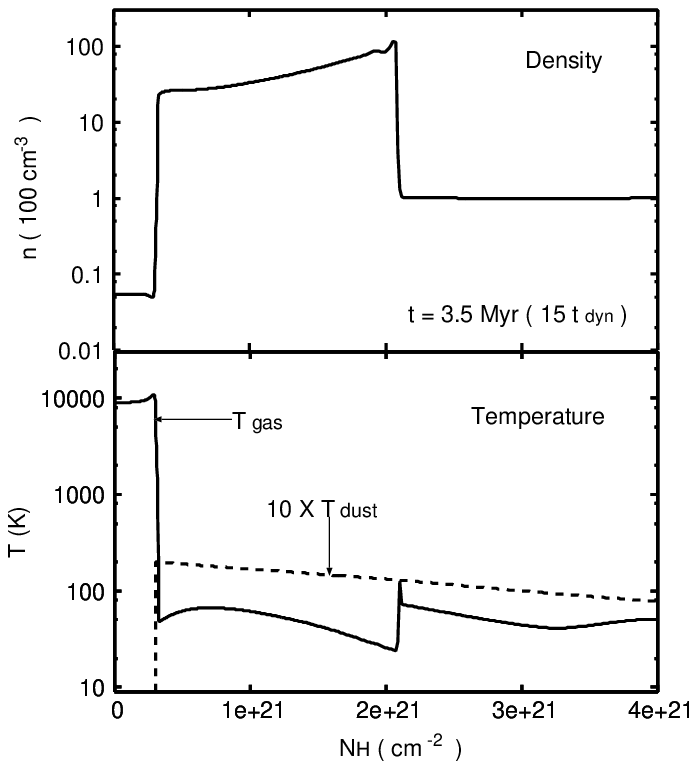}
\caption{
Same as Fig.\ref{fig:over_5000} but for model LD-S41.
These snapshots are at $t = 3.5$~Myr, which corresponds to 
$\sim 15~t_{\rm dyn}$.                                 
 }
\label{fig:over_50}
\end{center}
\end{figure}

The detailed structure of the shell differs significantly among
these models.
Fig.\ref{fig:over_5000} (Fig.\ref{fig:over_50})
shows the detailed structure of the density and the temperature 
within the shell at $t \sim 15~t_{\rm dyn}$ in model HD-S41
(LD-S41). 
These figures show that
the gas temperature in the inner region of the shell is higher 
in the model with the higher ambient density;
$T_{\rm HI} \sim 500$~K in model HD-S41 and
$T_{\rm HI} \sim 50$~K in model LD-S41. 
This is because the FUV flux at the IF, $F_{\rm FUV,i}$ is larger for 
the higher ambient density owing to the smaller Str\"omgren radius
(see eq.(\ref{eq:jfuvi_nsc})). 
The FUV radiation field, $G_{\rm FUV}$ defined by eq.(\ref{eq:fuvG}) at the IF 
is $G_{\rm FUV} \sim 10$ in LD-S41 and $\sim 5000$ in HD-S41 at the time 
step of both figures, $t \sim 15~t_{\rm dyn}$. 
Within the shell, the temperature has a negative gradient
and the density has a positive gradient, 
which is similar to model S41 (Paper I). 
Figs.\ref{fig:over_5000} and \ref{fig:over_50} show that
the changes in gas temperature and density
are milder in model LD-S41.
The gas temperature decreases from 500~K to 25~K through
the shell in model HD-S41 and from 50~K to 25~K in model LD-S41. 
These differences are also explained in equation (\ref{eq:sig_nsc})
and (\ref{eq:jfuvi_nsc}).
In model HD-S41, the FUV flux is strong and the column density
is high. The intense FUV radiation is shielded by the 
shell with the large column density. 
In model LD-S41, the incident FUV flux at the IF is
relatively weak and the column density of the shell is low. 
The gas temperature within the shell is lower and nearly
constant.
The FUV flux is attenuated from $G_{\rm FUV} \sim 5000$ to 0.1
through the shell in HD-S41, and from $G_{\rm FUV} \sim 10$ to 1.5 in
LD-S41 at the time of Fig.\ref{fig:over_5000} and \ref{fig:over_50}.
In model HD-S41, 
the gas temperature significantly decreases toward the outer part of the
shell, because the photoelectric heating becomes inefficient.
Conversely, the gas density of the outer 
region of the shell rises to $n \sim 10^6~{\rm cm}^{-3}$. 
The collisional heat transfer between the gas and dust is 
efficient there, and the gas temperature is almost equal to 
the dust temperature.
The the swept-up shell is geometrically thicker in model HD-S41.
Figs.\ref{fig:over_5000} and \ref{fig:over_50} also show that
the density jump at the IF is larger with the lower ambient density.
This can be explained by the analytic jump condition at the
weak D-type IF,
\begin{equation}
\frac{\rho_{\rm HII}}{\rho_{\rm HI}} 
= \frac{C_{\rm I}^2}{2 C_{\rm II}^2} (1 + \delta)
\sim
\frac{T_{\rm HI}}{T_{\rm HII}}
\end{equation}
\citep{MM84},
where $C_{\rm I}$ and $C_{\rm II}$ are the sound speed ahead
and behind the IF, and $\delta$ is the quantity less than unity.
Since the temperature $T_{\rm HI}$ is 
lower for the lower ambient density, the density at the inner 
edge of the shell is higher in model LD-S41.

%------------------------------------------%
\subsection{Fragmentation of the Shell}
\label{ssec:frg_n}
%------------------------------------------%

It is important to know whether or not the swept-up molecular layer
fragments, as the triggered star formation scenario predicts. 
We do not consider the dynamical instability, but only the
gravitational instability for the fragmentation of the shell, 
as discussed in Paper I.
In the lower panels of Figs.\ref{fig:fev_5000} and \ref{fig:fev_50},
the shaded region is the expected unstable region, 
where $t > (G \rho)^{-1/2}$. 
The swept-up shell in each model is expected to
suffer from  gravitational instability.
The unstable region spreads from the outer edge of the shell,
which is the same as in model S41 (Paper I).
In model HD-S41, the density is relatively low in the inner part
of the shell.
The unstable region appears at $t \sim 0.08$~Myr ($7.3~t_{\rm dyn}$)
and gradually extends from the cold molecular layer, 
as Fig.\ref{fig:fev_5000} shows.
As explained in the previous subsection, both H$_2$ and CO molecules
are shielded by the dust, and the shell is dominated by
these molecules when the shell becomes unstable.
In model LD-S41, the swept-up shell is relatively
cold and dense (see Fig.\ref{fig:over_50}). The unstable region 
rapidly spreads in the whole shell at $t \sim 1.5$~Myr ($6.5~t_{\rm dyn}$).
In this model, the shell is dominated by H$_2$ molecules
due to the efficient self-shielding effect by the time of the
shell-fragmentation.
However, the column density of the shell is still not high enough to
protect CO molecules by the dust extinction in the early expansion phase.
The accumulation of CO molecules begins after the shell becomes
unstable, which is different from model HD-S41.
Regardless of some differences, the expanding H~II region
can sweep up the molecular gas, and the fragmentation of the shell
is finally expected with different number densities.
The star formation will be triggered in the fragmented shell
along the lines of a ``collect and collapse'' scenario.

%%%%%%%%%%%%%%%%%%%%%%%%%%%%%%%%%%%%%%%%%%%%%
\section{Dependence on the Central Star}
\label{sec:sdep}
%%%%%%%%%%%%%%%%%%%%%%%%%%%%%%%%%%%%%%%%%%%%%%

%---------------------------------
\subsection{Models}
\label{ssec:mdl_s}
%---------------------------------

In this section, we investigate how the time evolution of
the H~II region, PDR, and the shell depends on the 
UV/FUV luminosity (mass) of the central star. 
For this purpose, we calculate the expansion around the
central star of $101.3 M_\odot$, $19.0 M_\odot$ and $11.7 M_\odot$ 
as representative cases. We use the same ambient number density, 
$n_{\rm H,0} \sim 10^3~{\rm cm}^{-3}$, as in model S41 (Paper I).
Model parameters of these calculated models are listed 
in Table 4.

%---------------------------------%
\subsection{Scaling Relations}
\label{ssec:scl_s}
%---------------------------------%

The less massive central star has smaller UV and FUV
luminosities, but a large luminosity ratio, $S_{\rm FUV}/S_{\rm UV}$ 
(see Table 1).
The Str\"omgren radius depends on the UV photon number luminosity,
$S_{\rm UV}$, as well as the number density, so that other relevant
quantities also depend on $S_{\rm UV}$. The scaling relations which are
similar to equation (\ref{eq:rst_nsc}) -- (\ref{eq:jfuvi_nsc})
are derived as,
\begin{equation}
\label{eq:rst_ssc}
R_{\rm st} \propto S_{\rm UV}^{1/3} \ ,
\end{equation}
\begin{equation}
\label{eq:tdy_ssc}
t_{\rm dyn} = R_{\rm st}/C_{\rm HII} \propto S_{\rm UV}^{1/3} \ ,
\end{equation}
\begin{equation}
\label{eq:msh_ssc}
M_{\rm sh} \propto n R_{\rm st}^3 \propto S_{\rm UV} \ ,
\end{equation}
\begin{equation}
\label{eq:sig_ssc}
N_{\rm sh} \propto n R_{\rm st} \propto S_{\rm UV}^{1/3} \ ,
\end{equation}
\begin{equation}
\label{eq:jfuvi_ssc}
F_{\rm FUV,i} \propto S_{\rm FUV}/R_{\rm st}^2 
              \propto (S_{\rm FUV}/S_{\rm UV})^{2/3} S_{\rm FUV}^{1/3},
\end{equation}
where $M_{\rm sh}$, $N_{\rm sh}$, and $F_{\rm FUV,i}$ are
the quantities at a given $t/t_{\rm dyn}$. 
As equation (\ref{eq:rst_ssc}) and (\ref{eq:tdy_ssc}) show,
the H~II region expands into the larger region on a longer timescale,
as the mass of the central star increases.
The FUV flux at the IF, $F_{\rm FUV,i}$ depends on both 
the luminosity ratio, $S_{\rm FUV}/S_{\rm UV}$ and 
the FUV luminosity, $S_{\rm FUV}$.
Since the dependence on the luminosity ratio is stronger 
than that on the FUV luminosity, 
the FUV flux, $F_{\rm FUV,i}$ becomes larger for the
lower mass central star. 
The calculated shell structure reflects
the column density, $N_{\rm sh}$ (eq.(\ref{eq:sig_ssc}))
and the FUV flux at the IF, $F_{\rm FUV,i}$, (eq.(\ref{eq:jfuvi_ssc}))
which is explained in \S\ref{ssec:rlt_s}.

%--------------------------------------------------%
\subsection{Results of the Numerical Calculation}
\label{ssec:rlt_s}
%--------------------------------------------------%

\begin{figure}[tb]
\begin{center}
\includegraphics[width=0.6\hsize]{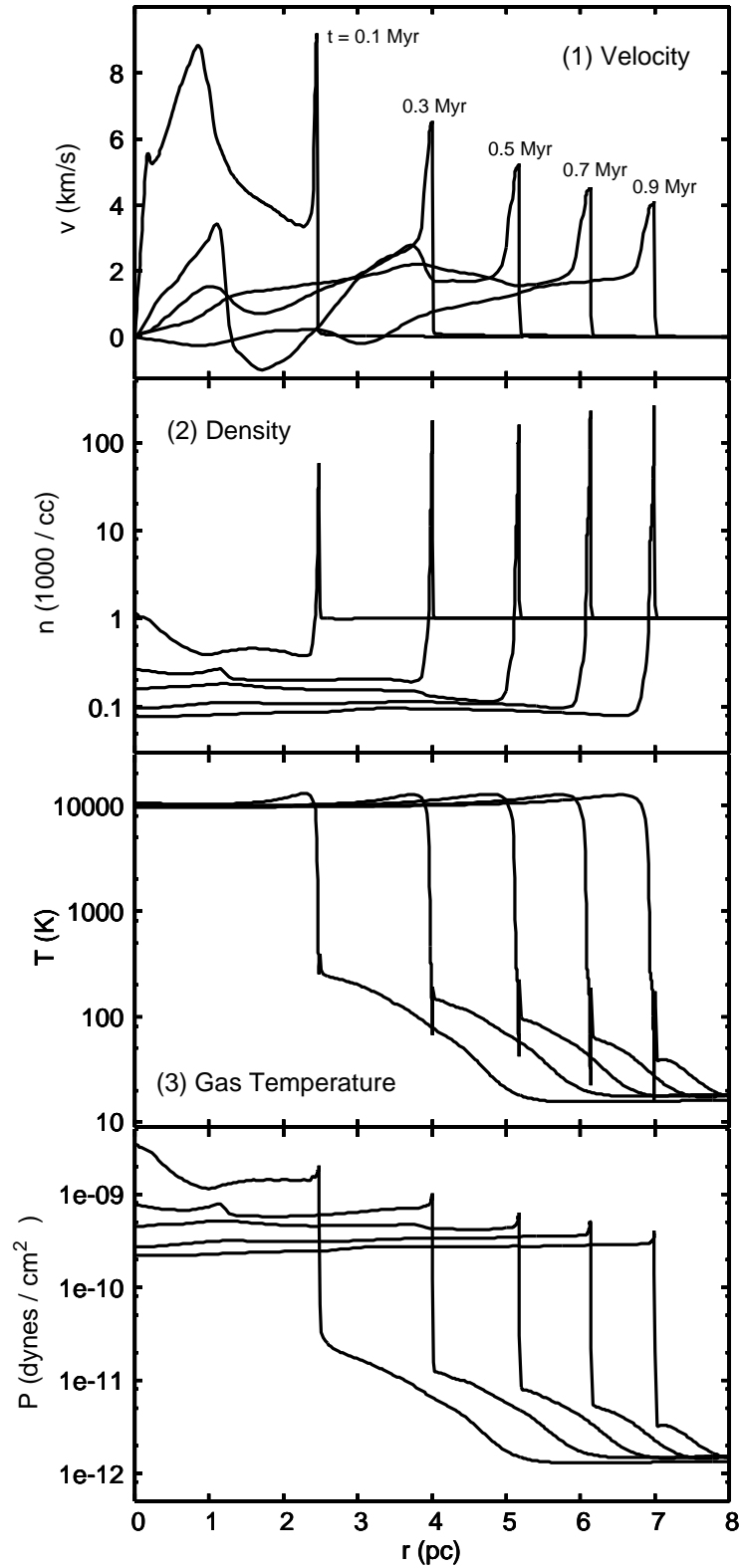}
\caption{Same as Fig\ref{fig:hev_5000} but for model S101. 
In each panel, five snapshots represent the profiles at 
$t = 0.1$, 0.3, 0.5, 0.7 and 0.9~Myr respectively.}
\label{fig:hev_s1}
\end{center}
\end{figure}

\begin{figure}[tb]
\begin{center}
\includegraphics[width=0.6\hsize]{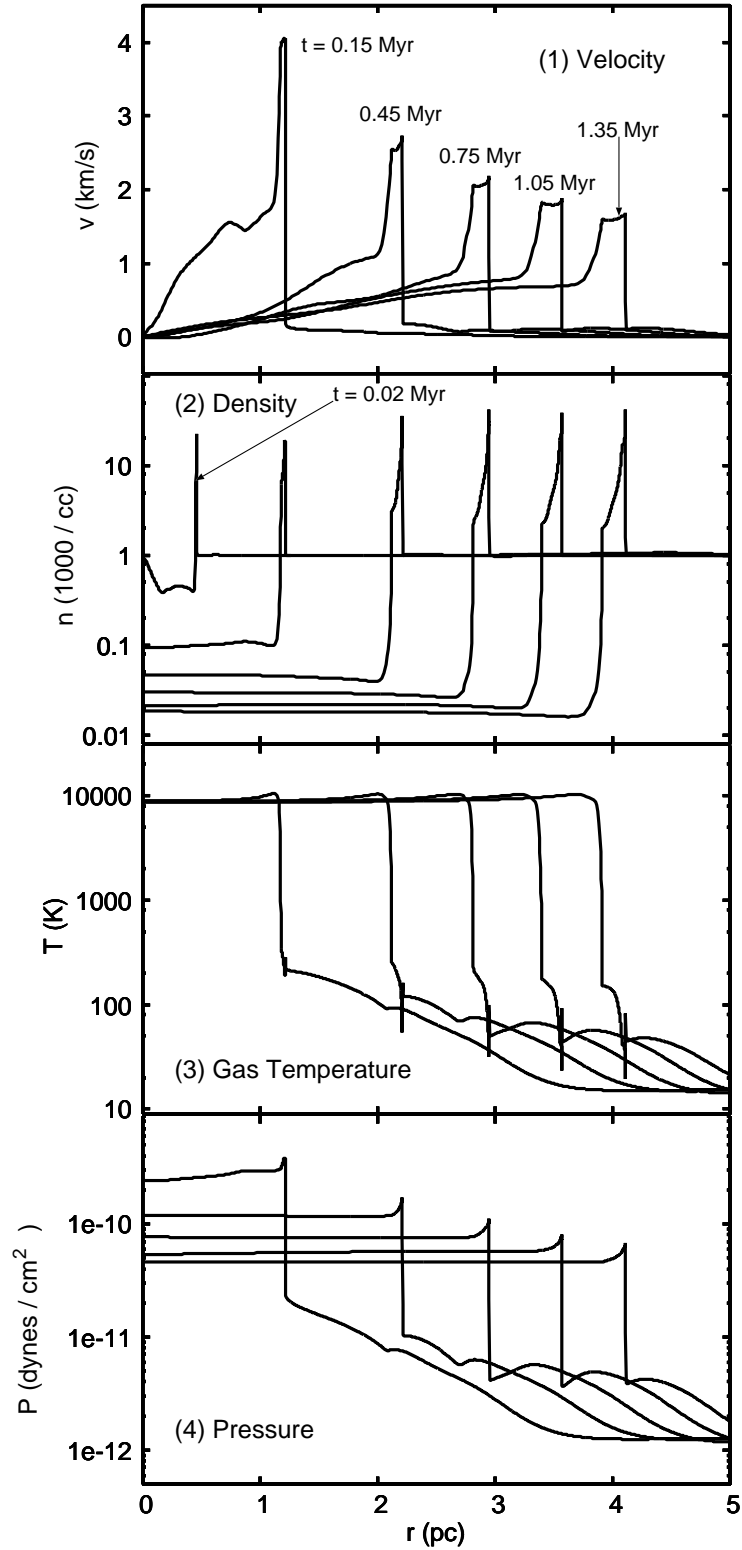}
\caption{Same as Fig\ref{fig:hev_5000} but for model S19. 
In each panel, five snapshots represent the profiles at 
$t = 0.15$, 0.45, 0.75, 1.05 and 1.35~Myr respectively.
We add one early snapshot of the density profile at $t = 0.02$~Myr.}
\label{fig:hev_s5}
\end{center}
\end{figure}

\begin{figure}[tb]
\begin{center}
\includegraphics[width=0.6\hsize]{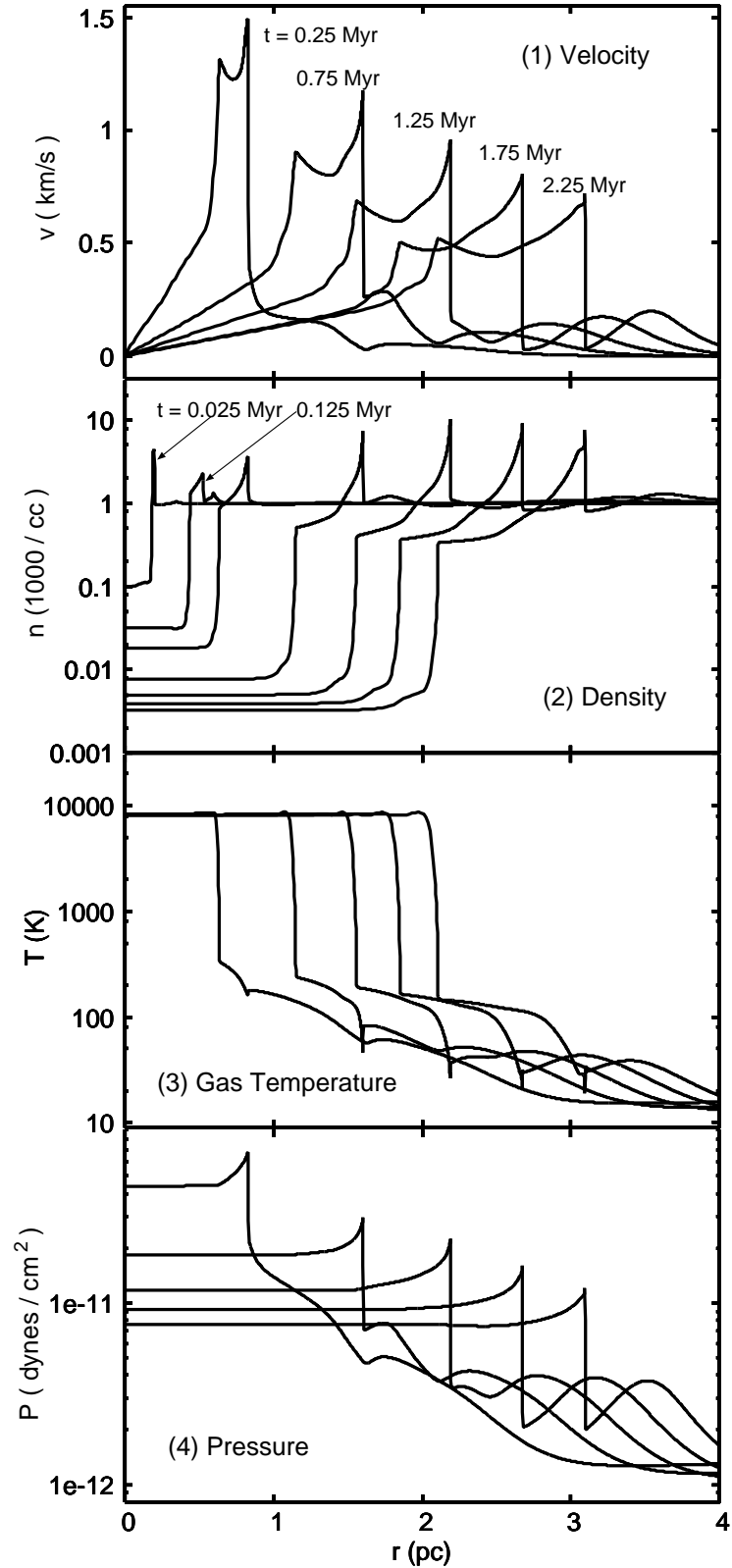}
\caption{
Same as Fig\ref{fig:hev_5000} but for model S12. 
In each panel, five snapshots represent the profiles at 
$t = 0.25$, 0.75, 1.25, 1.75 and 2.25~Myr respectively.
We add two early snapshots of the density profile at 
$t = 0.025$, and 0.125~Myr.
}
\label{fig:hev_s8}
\end{center}
\end{figure}

Figs.\ref{fig:hev_s1}, \ref{fig:hev_s5}, and \ref{fig:hev_s8} 
show the hydrodynamical
evolution of models S101, S19, and S12 respectively.
The basic time evolution features are similar among these
models. When the IF reaches the initial Str\"omgren radius, 
the SF emerges in front of the IF. 
In all these models, the H~II region expands into
the region of $\sim 1-10$~pc over a timescale of $\sim 1$~Myr.
However, the time normalized with $t_{\rm dyn}$ differs considerably
among these models, because  
$t_{\rm dyn}$ decreases as the UV luminosity decreases  
(eq. (\ref{eq:tdy_ssc})). 
For example, the time of 1~Myr corresponds to
$\sim 8~t_{\rm dyn}$ in model S101,
$\sim 40~t_{\rm dyn}$ in model S19, 
and $\sim 300~t_{\rm dyn}$ in model S12. 
Figs.\ref{fig:hev_s1} - \ref{fig:hev_s8} all show
that the PDR initially appears in front of the SF and is
finally taken into the shell over a timescale of $\sim 1-2$~Myr;
such evolution is largely different in terms of time normalized 
with $t_{\rm dyn}$.

\begin{figure}[tb]
\begin{center}
\includegraphics[width=0.8\hsize]{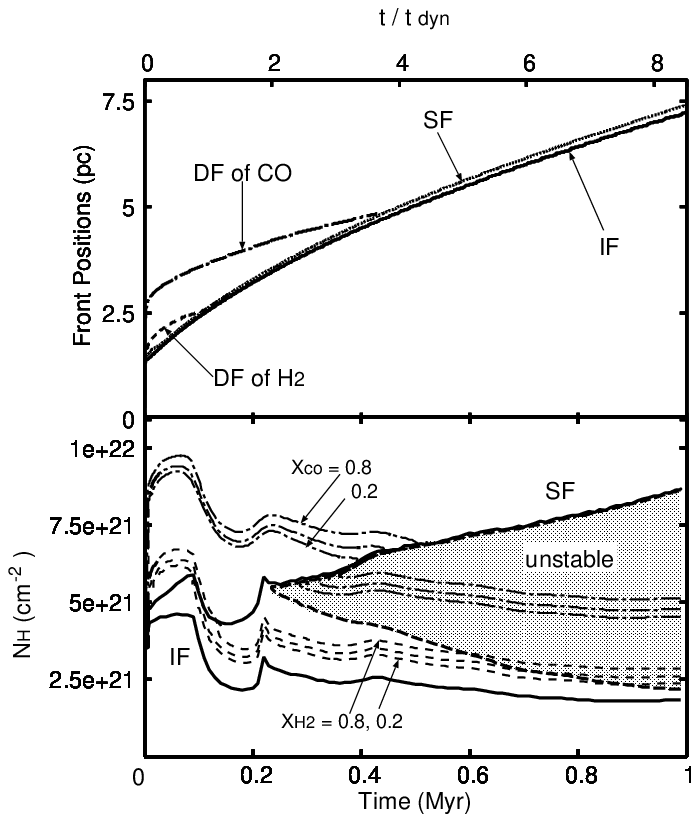}
\caption{
Same as Fig.\ref{fig:fev_5000} but for model S101.
}
\label{fig:fev_s1}
\end{center}
\end{figure}

\begin{figure}[tb]
\begin{center}
\includegraphics[width=0.8\hsize]{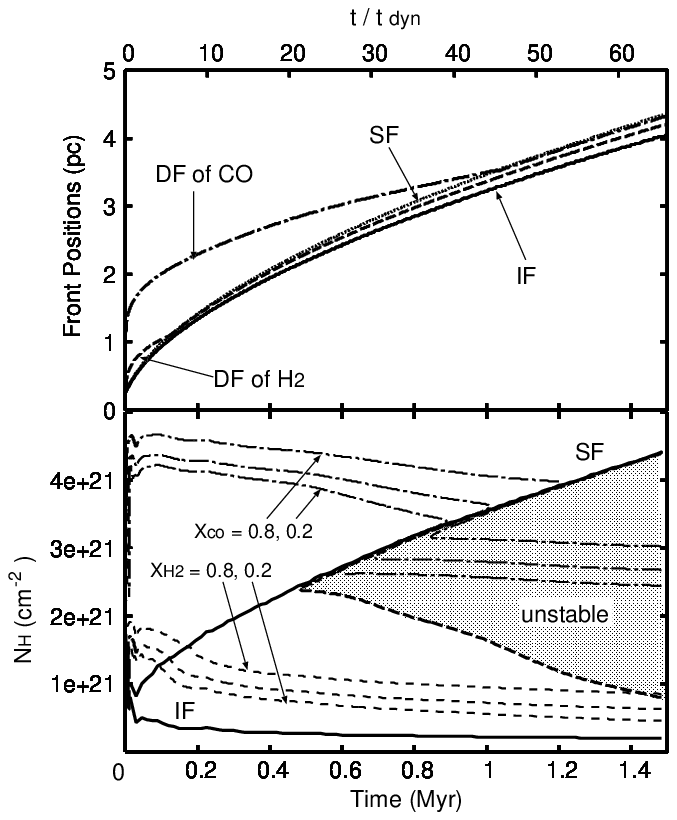}
\caption{
Same as Fig.\ref{fig:fev_5000} but for model S19.
}
\label{fig:fev_s5}
\end{center}
\end{figure}

\begin{figure}[tb]
\begin{center}
\includegraphics[width=0.8\hsize]{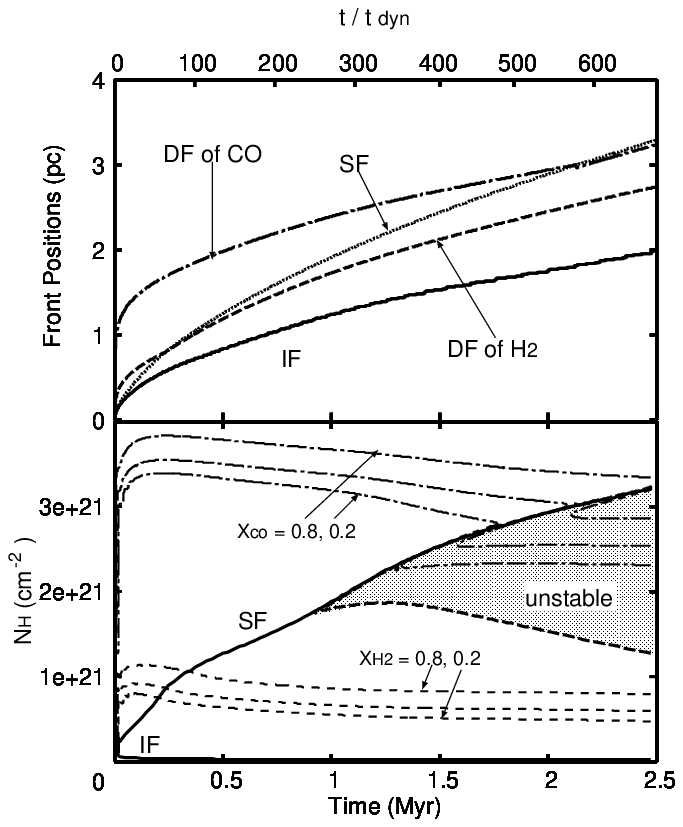}
\caption{
Same as Fig.\ref{fig:fev_5000} but for model S12.
}
\label{fig:fev_s8}
\end{center}
\end{figure}

\begin{figure}[p]
\begin{center}
\includegraphics[width=0.8\hsize]{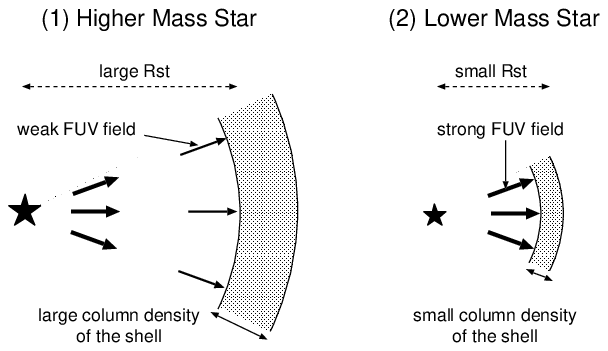}
\caption{Schematic figures for the H~II region and the swept-up
shell in models with different central stars.
These figures show the snapshot at one $t/t_{\rm dyn}$
in both models.
 }
\label{fig:sscale_schem}
\end{center}
\end{figure}

These differences are explained using equations (\ref{eq:sig_ssc})
and (\ref{eq:jfuvi_ssc}).
As equation (\ref{eq:jfuvi_ssc}) shows, the FUV flux at the IF
at a given $t/t_{\rm dyn}$, $F_{\rm FUV, i}$ is larger with the 
less massive star.
Conversely, the column density of the shell, $N_{\rm sh}$ 
is smaller for the less massive star, shown in equation (\ref{eq:sig_ssc}).
 At $t \sim 10~t_{\rm dyn}$, for example, $G_{\rm FUV} \sim 100$ 
at the IF and $N_{\rm sh} \sim 3 \times 10^{21}~{\rm cm}^{-2}$
in model S19, and $G_{\rm FUV} \sim 600$ and 
$N_{\rm sh} \sim 7 \times 10^{20}~{\rm cm}^{-2}$ in model S12.
Therefore, the strong FUV radiation is not shielded, and the reformed
molecules are easily photodissociated. 
Fig.\ref{fig:sscale_schem} is the schematic picture showing
the luminosity-dependence of $N_{\rm sh}$ and $F_{\rm FUV,i}$.
The accumulation
of molecular gas in the shell is significantly delayed on $t/t_{\rm
dyn}$.  The upper panels of Figs.\ref{fig:fev_s1} - \ref{fig:fev_s8}
show the time evolution of the various front positions in each model.
Both DFs of H$_2$ and CO molecules are taken into the
shell by the time of $4~t_{\rm dyn}$ in model S101,  
$45~t_{\rm dyn}$ in model S19,  and  
$550~t_{\rm dyn}$ in model S12. 
The times of the front
overtaking are totally different at $t/t_{\rm dyn}$, but these times
differ only by a factor of a few at time $t$; 
$t \sim 0.4$~Myr in model S101,
$t \sim 1$~Myr in model S19, and 
$\sim 2$~Myr in model S12. 
Eventually, sufficient molecular gas always gathers in the
outer region of the shell in a few Myrs with the
ambient number density of $n_{\rm H,0} \sim 10^3~{\rm cm}^{-3}$. 
The total mass of the swept-up shell (between the IF and the SF)
is $3.5 \times 10^4~M_\odot$, $9.8 \times 10^3~M_\odot$, 
and $3.8 \times 10^3~M_\odot$ in models S101, S19, 
and S12 respectively at the final time step
of Figs.\ref{fig:fev_s1} - \ref{fig:fev_s8}.
At that time step, more than 85\% of hydrogen atoms
are included in H$_2$ molecules in all models.
The fraction of CO molecules is 55\%, 40\%, and 20\% 
in each respective model.

\begin{figure}[tbp]
\begin{center}
\includegraphics[width=0.8\hsize]{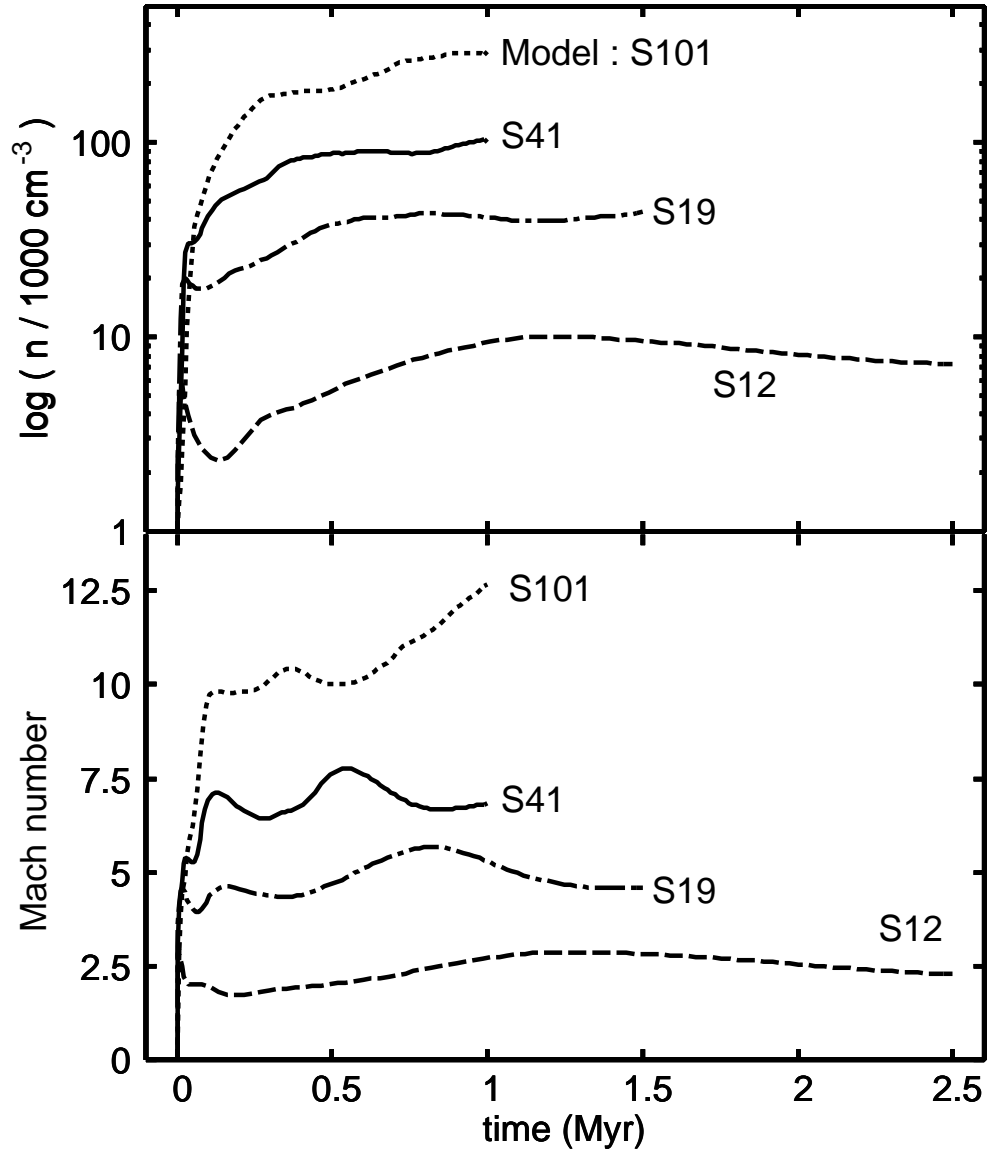}
\caption{{\it Upper panel : }Time evolution of the maximum density of 
the shell in model S101 (dotted line), S41 (solid line), S19 (dot-solid line) 
and S12 (broken line).
{\it Lower panel :} Time evolution of the Mach number of the SF
in each model. 
 }
\label{fig:star_dep}
\end{center}
\end{figure}

The upper panel of Fig.\ref{fig:star_dep} shows the time evolution 
of the maximum density of the shell in models S101, S41, S19 and S12. 
As this figure shows, the swept-up shell is
always denser with the more massive central star. 
This is explained with the time evolution of the velocity
of the shell.
The expansion of the H~II region approximately follows equation
(\ref{eq:exlaw}), and the expansion velocity decreases as, 
\begin{equation}
\dot{R}_{\rm IF}(t) = \sqrt{\frac43} 
                      C_{\rm HII} \left( 
                      1 + \frac74 \sqrt{ \frac43 } 
                                      \frac{t}{t_{\rm dyn}}
                           \right)^{-3/7} ,
\label{eq:exvel}
\end{equation}
because the gas pressure of the H~II region decreases. 
In model S101, the FUV radiation
is easily shielded because of the high column density of the shell.
The photoelectric heating outside the shell becomes inefficient, and 
the gas temperature in front of the SF quickly decreases.
At that time, $t/t_{\rm dyn}$ is still small, and the expansion 
velocity is still large according to equation (\ref{eq:exvel}).  
The Mach number of the SF is large, ${\cal M} \sim 10$. 
In model S12, it takes hundreds of $t_{\rm dyn}$ to shield 
the FUV radiation by the dust extinction through the shell. 
The expansion velocity significantly decreases by (\ref{eq:exvel}),
and the Mach number is only about a few.
This difference causes a different shell density 
in each model, because the density jump at the isothermal SF
is proportional to ${\cal M}^2$.
The lower panel of Fig.\ref{fig:star_dep} presents the time
evolution of the Mach number in each model.
This figure clearly shows that the Mach number is larger in the
model with the more massive central star.

During the expansion, the pressure gradient in the PDR outside
the shell hardly affects the hydrodynamics.
The pressure gradient outside the shell gradually disappears
as the FUV photons are shielded by the dust in the shell.
The SF does not emerge in front of the DFs in our calculations.

%------------------------------------------%
\subsection{Fragmentation of the Shell}
\label{ssec:frg_s}
%------------------------------------------%

Even with different central stars, the molecular gas is finally
accumulated in the swept-up shell. The lower panels of 
Figs.\ref{fig:fev_s1}, \ref{fig:fev_s5}, and \ref{fig:fev_s8} 
show the time evolution of the column density of each region 
in models S101, S19, and S12 respectively. 
In all these models, the number density of the
molecular layer finally becomes over ten times 
as high as the ambient density, and gravitational 
fragmentation is expected there. 
In Figs.\ref{fig:fev_s1} - \ref{fig:fev_s8}, the shaded region is
the unstable region expected in each model, where $t > (G \rho)^{-1/2}$.
As shown in these figures, the unstable region spreads in the shell,
as the DFs of CO are taken into the shell.
The unstable region appears at $t \sim 0.2$~Myr in model S101,
$t \sim 0.5$~Myr in model S19, and $t \sim 1.1$~Myr in model S12.
As explained in the previous subsections, 
these times are largely different
at $t/t_{\rm dyn}$; $\sim 2$ in model S101, 
$\sim 20$ in model S19, and $\sim 300$ in model S12. 
The FUV flux at the IF
is stronger with the less massive star, and it takes a long time
in $t_{\rm dyn}$ units to form the dense molecular layer 
shielding the FUV radiation field.

%%%%%%%%%%%%%%%%%%%%%%%%%%%%
%%%%%%%%%%%%%%%%%%%%%%%%%%%%
\section{Discussions}
\label{sec:dcs}
%%%%%%%%%%%%%%%%%%%%%%%%%%%%
%%%%%%%%%%%%%%%%%%%%%%%%%%%%

%------------------------------------------------%
\subsection{Dust Grains} 
\label{ssec:dust}
%------------------------------------------------%

\subsubsection{Dust Grains in H~II Region}
\label{ssec:dusthii}

In the above calculations,
we have not included the dust grains in the H~II regions
to avoid uncertainties (e.g., size distribution), but some 
observations indicate their presence \citep[e.g.,see][]{Sp78}.
The dust absorption of the UV and FUV radiation will decelerate
the propagation of the IF and DFs and affect the structure of the 
swept-up shell. 
\citet{WD01} have noted that the photoelectric heating 
can be as important as the photoionization heating, even in the 
H~II region.
The importance of the dust absorption in the H~II region
is often evaluated with the column density of the dust-free
static H~II region,   
\begin{equation}
N_{\rm HII} \propto n R_{\rm st} \propto n^{1/3} S_{\rm UV}^{1/3}  
\label{eq:nhii}
\end{equation}
\citep{Pt72, Ar04}, 
where $R_{\rm st}$ is the dust-free Str\"omgren radius, $n$ is
the ambient number density, and $S_{\rm UV}$ is the UV photon
number luminosity of the central star.  
The radius of the corresponding dusty H~II region is smaller for
larger $N_{\rm HII}$. \citet{Ar04} has also shown that the dust 
absorption in the H~II region gradually becomes inefficient, 
as the H~II region dynamically expands.

In order to evaluate the effect of the dust in the H~II region, 
we have calculated some models including the dust
in the H~II region as well as the outer PDR.
In models S41-HD and S41-LD, the radius of the H~II region and PDR
decreases only by 10\% and 3\% compared with the dust-free counterparts.
In models S101 and S41, the reduction of each region is
20\% and 5\% of the dust-free radius. 
These results agree with equation (\ref{eq:nhii}),
which shows that the dust absorption in the H~II region 
becomes significant with the higher ambient number density, 
or the more massive central star.
In all of the above calculations, however, the dust absorption
in the H~II region is not so significant.
Furthermore, the dust in the H~II region also absorbs FUV photons
in addition to  UV photons. As briefly mentioned in Paper I,  
this only slightly promotes the accumulation of molecules in the shell.
Therefore, dust grains in the H~II region hardly affect our 
conclusions. The dust absorption in the H~II region is more 
important in the denser and more compact region 
(e.g., ultra compact H~II region),
where the detailed treatment of grains is required.

\subsubsection{Reformation of Hydrogen Molecules on Grain Surface}

The reformation of H$_2$ molecules on the grain surface 
is one of the key processes in our calculations.
The FUV radiation is consumed in photodissociating the reformed
molecules in the shell, which accelerates the accumulation of 
H$_2$ molecules.
However, there still remain uncertainties in the reformation rate, 
which depends on the detailed behavior of H atoms and H$_2$ 
molecules on the grain surface.
At high dust temperatures, for example, H atoms absorbed on the 
grain surface evaporate before the reformation of H$_2$ molecules.
The critical temperature, $T_{\rm cr}$, is about 20~K on pure grains,
but \citet{HS71} have shown that the irregular surface with impurity
sites raises $T_{\rm cr}$ to $25~{\rm K} < T_{\rm cr} < 50~{\rm K}$.
Recently, \citet{CT02, CT04} have refined the absorption process
of H atoms on the grain surface, and obtained a much 
higher critical temperature of $T_{\rm cr} \sim 100$~K.
In our calculations presented in \S4 and 5, 
we have not included the high-temperature cut-off 
in the reformation rate, for simplicity. 
We analyze the validity of this treatment here.

We calculate some models including the 
cut-off using the functional form given by \citet{HM79} (HM79). 
In each model, two cases of $T_{\rm cr} = 65$~K (HM79) and 
35~K are applied. 
With $T_{\rm cr} = 65$~K, however, the H$_2$ molecular
abundance in the shell is hardly affected in all examined models. 
With $T_{\rm cr} = 35$~K, the accumulation of H$_2$ molecules
in the shell is slightly delayed, but the shell is finally dominated
by H$_2$ molecules.
The H$_2$ abundance in the shell is reduced only by 2\% in model 
S41 (Paper I) and 5\% in model S41-HD at $t \sim 20~t_{\rm dyn}$.
In the region where H$_2$ molecules accumulate, 
the FUV radiation is well shielded and the dust temperature
is low. Therefore, even if the reformation is inefficient
above several ten K, our conclusions do not change.

\subsubsection{Small Grains}

Our calculations do not include detailed energy and ionization 
balance of the small grains such as polycyclic aromatic hydrocarbons 
(PAHs). The ionization rate and size distribution of small grains 
are important factors for the PDR structure. 
The photoelectric heating rate depends on the abundance and 
ionization rate of PAHs, and the dust attenuation law of FUV photons 
may be slightly affected by the change of the grain size
distribution. We have calculated some models with different 
photoelectric heating rates given by \citet{WD01} 
based on the different abundance of small grains. The temperature 
profile in the PDR would be modified by a factor of less than 2, but 
this does not affect our conclusions. Some recent 
observations detect the strong PAH emission around the H~II region. 
This emission sometimes shows the clear ring-like structure, 
which should trace the shell \citep[e.g.,][]{Zv06}.
The expected PAH emission related to the time evolution of the shell
will be very useful for these observations, which should be included
in future modeling.

%---------------------------------------------------%
\subsection{Realistic Structure of Molecular Clouds}
%---------------------------------------------------%

The actual molecular clouds have complex
structures rather than the homogeneous density distribution
assumed in this paper. Below, we briefly discuss how 
realistic density structures affect the time evolution of the
swept-up shell, PDR, as well as the H~II region.

\subsubsection{Density Gradient}

When the H~II region forms near the edge of a molecular cloud, or
in a cloud with the density gradient (e.g., $n \propto r^{-w}$), 
its time evolution is different from that in a homogeneous
medium. The high-pressure ionized gas is hardly trapped, and
flows out of the cloud \citep[][]{Tt79, FTB90}.
Such ``blister-like'' or ``champagne flow'' features
have been observed in some H~II regions.
In these regions, only a small fraction of the swept-up gas remains 
in the shell, and the expanding IF finally destroys the parental 
molecular cloud. As argued by \citet{Wt79} and \citet{FST94},
the rapid ionization of the molecular gas will quench the
star formation activity in the clouds.
Furthermore, we note that the star-forming capacity of the clouds
will be limited by FUV photons as well as UV photons.
Unless the swept-up gas remains in the shell, 
stellar FUV radiation is hardly attenuated through the shell.
Such unshielded FUV photons dissociate molecules efficiently
over a much wider region than the H~II region, 
which reduces the star formation efficiency of the clouds.  
The dynamical expansion of the H~II region, PDR, and the
shell in a cloud with a density gradient has not been 
studied well, and this will be explored in our forthcoming papers.

\subsubsection{Clumpiness}

Real molecular clouds generally have clumpy or 
hierarchical structures. The pre-existing density inhomogeneity
in molecular clouds can modify the structure and the
evolution of the H~II region, PDR, and the swept-up shell.

In the H~II region, the pre-existing clumps are gradually
photoevaporated by stellar UV radiation.
These clumps are basically smoothed 
out in the sound crossing time of each clump, which is much shorter 
than the dynamical time of the H~II region, 
$t_{\rm dyn} = R_{\rm st}/C_{\rm HII}$.

Outside the H~II region, the PDR structure depends on how deep the 
FUV radiation penetrates into the molecular region.
Some previous work \citep[e.g.,][]{Bs90} has shown that
the radiative transfer in a clumpy medium can be 
reduced to that of a homogeneous medium with the same average
density and lower effective extinction.
In the PDR around the H~II region, the FUV photons will easily 
seep into the molecular region, threading the clumpy structure.
This will delay the SF and IF overtaking the DFs, and the
accumulation of molecules in the shell.
In clumpy molecular clouds, on the other hand, 
most of the gas is thought to be contained in optically thick clumps.
Molecules in these opaque clumps will be protected against the FUV
radiation by the self-shielding of these clumps.
In order to study these two competing effects, multi-dimensional and
time-dependent calculations are needed.
Note, however, that our 1-D calculation has successfully
explained the observed molecular abundance around Sh104 (Paper I).
Thus we expect that our 1-D calculations should be a good approximation
for the averaged physical quantities.

\subsubsection{Turbulent Velocity Field}

Observation shows that there are the supersonic turbulent
motions in molecular clouds \citep[e.g.,][]{ES04}. 
Since the velocity dispersion 
increases with the lengthscale ($\sim$ a few km/s at about 10~pc), 
the effect becomes important with the lower ambient density.
The expanding shell actually dissolves when the shell velocity
becomes less than the velocity dispersion of the turbulent motion.
In \S5, we have shown that the shell velocity becomes less than
a few km/s when the molecular gas accumulates in the shell 
with the lower-mass star.
In this case, the shell will dissolve without gathering the molecular
material, and the triggered star formation will not occur. 
We study the net feedback effect in the turbulent molecular cloud
in a subsequent paper.

%-----------------------------------------%
\subsection{After the Cloud Destruction}
%-----------------------------------------%

In this paper, we have considered only the expanding H~II region 
in the molecular cloud, which can trigger the propagation of  
star formation activity within the cloud.
After the molecular cloud disperses due to the negative feedback
from massive stars and/or the external turbulent motion, stellar
UV photons spread into the diffuse neutral medium.
The H~II regions expand in the ambient neutral medium, 
but the SF preceding the IF should also compress the neutral gas 
to enable the rapid reformation of molecules 
\citep[e.g., ][and references therein]{ki00, Bg04}.
The role of the giant H~II regions expanding in the diffuse neutral
medium will be explored in our next paper, 
Hosokawa \& Inutsuka (2006, in preparation).

%%%%%%%%%%%%%%%%%%%%%%%%%%%%
%%%%%%%%%%%%%%%%%%%%%%%%%%%%
\section{Conclusions}
%%%%%%%%%%%%%%%%%%%%%%%%%%%%
%%%%%%%%%%%%%%%%%%%%%%%%%%%%

In this paper, we have studied the time evolution of the H~II region,
PDR, and the swept-up shell around a massive star, performing
the numerical calculation of radiation-hydrodynamics.
Following Paper I, 
we have investigated how the time evolution changes with the different
ambient number density and the luminosity (mass) of the central star.
The basic evolution with different parameters is qualitatively
similar to the fiducial one explained in Paper I, but
quantitatively different.
We have derived some simple scaling relations, 
(\ref{eq:rst_nsc})-(\ref{eq:jfuvi_nsc}) and 
(\ref{eq:rst_ssc})-(\ref{eq:jfuvi_ssc}),
with which we can well understand what causes the differences among models.

First, we have analyzed the dependence on the ambient number
density. Our results are summarized as follows:
\begin{itemize}
\item[1.] 
At the typical ambient number density of GMCs, 
$n_{\rm H,0} = 10^{2-4}~{\rm cm}^{-3}$, the molecular gas
is finally accumulated in the shell, and gravitational
fragmentation is expected. 
\item[2.] 
The timescale of the evolution, size of the H~II region, and the
swept-up mass in the shell quantitatively change with the different 
ambient number densities.
With the lower number density, the H~II region expands into
the larger region over a long timescale, 
so that the swept-up mass in the shell is larger.
The expected timescale within which the layer becomes gravitationally
unstable is also longer with the lower ambient number density.  
\item[3.]
The H$_2$ molecules rapidly accumulate in the shell,
but CO molecules are more sensitive to the shielding effect of FUV photons.
With the lower number density, the accumulation of CO molecules
in the shell is delayed owing to the small column density of the shell.
At $n_{\rm H,0} = 10^{4}~{\rm cm}^{-3}$, the shell is dominated
by CO molecules before the fragmentation.  
At $n_{\rm H,0} = 10^{2}~{\rm cm}^{-3}$, however, the
shell-fragmentation is expected to take place before the 
accumulation of CO molecules.
If the shell fragments into clumps with high column density,
CO molecules will be formed there owing to the self-shielding of clumps. 
\end{itemize}
Next, we have analyzed the dependence on the luminosity (mass)
of the central stars. 
Our results are the following:
\begin{itemize}
\item[4.]
With the central stars of 
$M_{\rm *} = 101-12~M_\odot$, the molecular gas
is finally accumulated in the shell, and gravitational
fragmentation is expected.
\item[5.] 
The timescale of the shell-fragmentation is relatively insensitive
to the luminosity of the central star.
At $n_{\rm H,0} \sim 10^3~{\rm cm}^{-3}$,
the unstable region spreads in the shell over $0.5-2$~Myrs with
the central star of $M_{\rm *} = 101-12~M_\odot$, whose
UV and FUV luminosities differ by 2-4 orders of magnitude.
\item[6.]
The maximum density of the shell is always higher with the more massive
central star. 
This is due to the difference in the Mach number of the SF.
With the higher-mass star, the FUV radiation is easily shielded
and the temperature in front of the SF quickly decreases, which
raises the Mach number. 
\end{itemize}
According to the above results, we can conclude that an expanding 
H~II region in homogeneous molecular gas should be an efficient trigger
for star formation. 
Once the massive star is ignited in the cloud, the star formation 
can propagate to the remaining molecular material around the
exciting massive star.

{\acknowledgements 
We are grateful to Shin Mineshige, Takashi Nakamura, Toru Tsuribe,
Kazu Omukai, Hiroshi Koyama, and Akira Mizuta for their useful 
comments and encouragement.
We also owe thanks to the YITP and NAOJ computer systems
for the numerical calculations.
SI is supported by the Grant-in-Aid  (15740118, 16077202)
from the Ministry of Education, Culture, Sports, Science, and
Technology (MEXT) of Japan.}

\clearpage
\appendix

%%%%%%%%%%%%%%%%%%%%%%%%%%%%%%%%%%%
\section{Finite Difference Method}
%%%%%%%%%%%%%%%%%%%%%%%%%%%%%%%%%%%

In this section, we explain the difference method for
the ionization equation and the energy equation, (\ref{eq:egy_ite}).
Other rate equations are differenced in the same way as the ionization
equation.
For simplicity, we temporarily omit all indices and 
drop the frequency-dependence of the radiation field in the ionization
equation,
\begin{equation}
\label{eq:ion_simple}
\frac{dX}{dt} = (1-X) \sigma F + R \ ,
\end{equation}
where $R$ means some processes other than photoionization
(e.g., recombination etc).
The simplest difference form of (\ref{eq:ion_simple}) is,
\begin{equation}
\label{eq:ion_s_dif}
\frac{X_j^{~n+1} - X_j^{~n}}{\Delta t} = \sigma 
                      \left\{ (1 - X_j) F_j \right\}^{n+1/2}
                    + R_j^{~n+1/2},
\end{equation}
where all physical quantities are defined at the cell center. 
The time average, $Q_j^{~n+1/2}$ for a quantity, $Q$ is defined as
\begin{equation}
Q_j^{~n+1/2} = \frac{ Q_j^{~n+1} + Q_j^{~n}}2 \ .
\end{equation}
Using the radiative transfer equation, (\ref{eq:juv}),
the UV photon number flux in the $j$-th cell, $F_j$ is written in
time steps,
\begin{equation}
\label{eq:j_simple}
F_{j} = F_{j-1} \frac{r_{j-1}^2}{r_j^2}
                \exp \left( 
                   -  \sigma \int_{r_{j-1}}^{r_{j+1}}
                              n (1-X)~dr                       
                     \right)_.
\end{equation}
Eliminating $F_{j}^{~n+1}$ in equation (\ref{eq:ion_s_dif})
with (\ref{eq:j_simple}), we can calculate $X_{j}^{~n+1}$ 
iteratively with other rate equations and the energy equation.
However, this differenced form provides no reliable results.
If the optical depth through the $j$-th cell is thick enough, 
$\Delta \tau_j \gg 1$, 
the UV flux at the $j$-th cell center, $F_j$, significantly decreases 
from the UV flux at the previous cell center, $F_{j-1}$. 
Even if a sufficient number of UV photons reaches the cell interface at
$r_{j-1/2}$, only a few photons may reach the $j$-th cell center and 
the $j$-th cell will not be photoionized. 
This situation occurs at the ionization front (IF),
where the optical depth changes sharply.

The quick way to improve the scheme is to define the flux
at the inner cell interface, not in the cell center.
However, we adopt another more sophisticated differenced form 
for rate equations.
With equation (\ref{eq:ion_simple}) and radiative transfer equation
(\ref{eq:uvtr}), the ionization equation becomes,
\begin{equation}
\label{eq:rate_dif_1}
\frac{d X}{dt} = - \frac{1}{n r^2}
                             \frac{\partial ( r^2 F )}{\partial r}
                           + R \ .
\end{equation}
First, equation (\ref{eq:rate_dif_1}) is differenced for the spatial
grids,
\begin{equation}
\label{eq:rate_dif_2}
\frac{dX_j}{dt} = 
\frac{r_{j-1/2}^2 F_{j-1/2} 
      [ 1 - \exp(-n_j \sigma (1-X_j) \Delta r_j) ]}
     { n_j ( r_{j+1/2}^3 - r_{j-1/2}^3 )/3 } + R_j \ .
\end{equation}
In the right-hand side, the first term represents the consumption
rate of UV photons for the photoionization per unit volume. 
This equation assures that
the number of ionized atoms per unit time is equal to that of 
the consumed photons per unit time in the $j$-th cell, 
if other processes, $R_j$, are neglected. 

Since the ionization rate sharply changes at the IF,
$X_j$ should be calculated with some precision 
\citep[e..g,][]{Mt65, Tt76}. 
We approximate the time evolution of
the ionization rate between two time steps with the power-law form,
\begin{equation}
\label{eq:plaw}
X_j (t) = X_j^{~n} + \frac{ X_j^{~n+1} - X_j^{~n}}{\Delta t} t \ , 
\qquad (0 \leq t \leq \Delta t)
\end{equation}
and analytically integrate (\ref{eq:rate_dif_2}) by time, 
$t$ from $t^n$ to $t^{n+1}$. 
Other terms are calculated in the same manner.
With the frequency-dependence of the radiation field and the dust
extinction, we can derive the differenced form in a similar way. 
We use the same differenced form to calculate the photodissociation
rate of H$_2$ and CO molecules.
The self-shielding effect works with the low column density ;
$N_{\rm H_2} > 10^{14}~{\rm cm}^{-2}$ for H$_2$ molecules 
\citep{DB96} and $N_{\rm CO} > 10^{15}~{\rm cm}^{-2}$ for CO 
molecules \citep{Le96}. 
The column density of one cell can be much higher than these values
in our time-dependent calculations,
and we do not use the shielding function given by \citet{DB96} and
\citet{Le96}. 

The energy equation (\ref{eq:egy_ite}) is differenced as, 
\begin{eqnarray}
\label{eq:egy_dif}
\frac{T_j^{~n+1} - T_j^{~n}}{\Delta t} &+& 
\left( \frac{T}{2w} \right)^{n+1/2} 
\left(  \frac{ {X_{\rm H,j}}^{n+1} - {X_{\rm H,j}}^{n}}{\Delta t} 
    + 3 \frac{ {X_{\rm H^+,j}}^{n+1} - {X_{\rm H^+,j}}^{n}}{\Delta t} 
 \right)  \nonumber \\
&+& \frac{\mu m_{\rm H} (\gamma -1)}{k}
    \left\{ \left( \frac{u}{w} \right)^{n+1/2} 
                 r^2 \frac{\partial p}{\partial m} 
         -  \left( \frac{1}{w} \right)^{n+1/2} 
                 \frac{\partial (r^2 u p)}{\partial m} 
         -  \left( \frac{1}{w} \right)^{n+1/2}
                 \frac{\partial (r^2 q_{\rm cond})}{\partial m}
    \right\} \nonumber \\
&=& \frac{\gamma - 1}{k} 
    \left( \frac{\Gamma - \Lambda}{w} \right)^{n+1/2}.
\end{eqnarray}
With differenced ionization equation (\ref{eq:rate_dif_2}), 
other similar rate equations, and the energy equation
(\ref{eq:egy_dif}), we iteratively solve the new temperature,
$T_j^{~n+1}$ and chemical compositions, ${X_{s,j}}^{n+1}$.

%%%%%%%%%%%%%%%%%%%%%%%%%%%%%%%%%%%%%%%%%%%%%%%%%%%%%%%%%%%%%%%%%%%
\section{Necessity of Thermal Conduction -- Field Condition --} 
\label{sec:field}
%%%%%%%%%%%%%%%%%%%%%%%%%%%%%%%%%%%%%%%%%%%%%%%%%%%%%%%%%%%%%%%%%%%

We include the heat transfer by thermal conduction
in the energy equation (\ref{eq:egyeq}). 
Recently, \citet{ki04} have shown
that the thermal conduction must be included 
to calculate the gas-dynamics with the thermal instability (TI). 
The conduction stabilizes the
TI for the short wavelength and the critical wavelength is called
the Field length, 
\begin{equation}
\lambda_{\rm F} = \left( \frac{KT}{n \Lambda} \right)^{1/2},
\end{equation}
\citep{Fd65} where $K$ is the conductivity and $\Lambda$ is the cooling rate
per hydrogen atom. \citet{ki04} have shown that the Field length
must be resolved with several spatial cells to converge the
calculation, increasing the number of spatial grids.
In our calculations, the TI occurs around 
$T \sim {\rm several} \times 10^3$~K mainly
at the IF. If we do not include 
the thermal conduction, unphysical oscillation spreads from 
the IF and the calculation does not converge.
Therefore, we need to include the thermal conduction in the energy equation.

Since the Field length is generally much shorter than the characteristic
length scale (e.g., Str\"omgren radius) in our calculations,
we adopt another available method to resolve Field length.
We first determine the width of the cell, 
$\Delta r_j$ and calculate the conductivity, $K_j$
so that the Field length is longer than the given width of the
cell. The index $j$ means the difference step in space. 
If we set $\lambda_{\rm F} \sim 3 \Delta r_j$,
such conductivity is calculated as,
\begin{equation}
\label{eq:cduc}
K_j = \frac{9 n_j \Lambda_j}{T_j}~(\Delta r_j)^2 .
\end{equation}
The conductivity, $K_j$ is calculated every time step with equation
(\ref{eq:cduc}).
If we set 100 cells per initial Str\"omgren radius for the  
ambient number density, $10^3~{\rm cm}^{-3}$, the calculated conductivity
is, $K \sim 4 \times 10^{11-10}~{\rm erg}~{\rm cm^{-1}}
~{\rm K^{-1}}~{\rm s^{-1}}$, which is significantly larger than the 
standard value.
However, the conduction with equation (\ref{eq:cduc})
smooths out the temperature structure only on the Field length scale, 
which covers a few cells 
($\lambda_{\rm F} \sim 3 \Delta r_j$).  
Since the meaningful temperature profile changes over a much 
longer length scale,  
we use the equation (\ref{eq:cduc}) to calculate the
conductivity.

%%%%%%%%%%%%%%%%%%%%%%%%%%%%%%%%%%%%%%%%%%%%%%%%%%%%%%%%%%%%
\section{Doppler Shift of Lines} 
\label{sec:dopp}
%%%%%%%%%%%%%%%%%%%%%%%%%%%%%%%%%%%%%%%%%%%%%%%%%%%%%%%%%%%%

\subsection{Doppler Effect for Cooling Lines}

The dominant cooling processes in the PDR are the radiative
loss by the fine-structure transitions, [OI] 63.1$~{\rm \mu m}$ and
[CII] 157.7$~{\rm \mu m}$ \citep{HT99}.
These lines can be optically thick at
$N_{\rm H} > 10^{21}~{\rm cm}^{-2}$ \citep{HM79, HM89}, 
and we use the escape probability for these lines \citep{dJ80, TH85}.
Since the swept-up shell expands at a supersonic 
velocity relative to the ambient gas, a few - 10~{\rm km/s},
we take account of the Doppler shift for the trapping effect. 
The gas velocity in the shell is usually almost constant, 
and we separately calculate the escape probability for the
gas in the shell and ahead of the shell.
For the gas in the shell, photons can escape from the inner
(IF) and outer edge (SF) of the shell, 
\begin{equation}
\beta_{\rm esc} =  f_{\rm esc}(\tau_{x,{\rm s1}})
                 + f_{\rm esc}(\tau_{x,{\rm s2}}) ,
\end{equation}
where $\beta_{\rm esc}$ is the escape probability, 
$\tau_{x,{\rm s1}}$ and $\tau_{x,{\rm s2}}$ are the optical depth
averaged over the line to the IF and the SF respectively,
and $f_{\rm esc}(\tau)$ is defined as,
\begin{eqnarray}
f_{\rm esc}(\tau) &=&  \frac{1 - \exp(-2.34 \tau)}{4.68 \tau}
                       \qquad \qquad \tau < 7 \ , \nonumber \\
   &=&   \left\{ 
          4 \tau \left[ \ln \left( \frac{\tau}{\sqrt{\pi}} \right) 
                  \right]^{0.5}
         \right\}^{-1}        
         \ \ \ \  \tau \geq 7  
\end{eqnarray}
\citep{dJ80}. For the gas ahead of the shell, we assume that 
only half of the photons can escape from the SF,
\begin{equation}
\beta_{\rm esc} = f_{\rm esc}(\tau_{x,{\rm ad}}) ,
\end{equation}
where $\tau_{x,{\rm ad}}$ is the optical depth to the SF. 
The CO molecules are another important coolant in the PDR.
We do not solve the non-LTE rotational level population, 
to save the computational time, but adopt
the analytical formulae given by \citet{HM79} including the
trapping effect.
We take account of the Doppler shift for the trapping effect
in the same manner as the fine-structure lines. 
In our calculations, the rovibrational lines of H$_2$ molecules
are reliably optically thin.

\subsection{Doppler Effect for Photodissociating Lines}

The FUV photons emitted from the central star photodissociate
the molecules, both in the swept-up shell and in the unshocked ambient
medium. However, we do not include the Doppler effect for the 
photodissociating lines of H$_2$ and CO molecules. 
Below, we briefly verify this approximation.
The Doppler shift in the FUV energy range is, 
$h \Delta \nu_{\rm d} = h \nu_0 (u_{\rm sl}/c) \sim 10^{-4}~{\rm eV}$, 
where $h \nu_0 \sim 10~{\rm eV}$ is the characteristic energy 
of FUV photons, $u_{\rm sl} \sim 10~{\rm km/s}$ is the typical
velocity of the shell, and $c$ is the speed of light.
The Doppler shift is larger than the Doppler width of lines
owing to the supersonic velocity of the shell.
As noted in \S\ref{ssec:beq}, however, 
the absorption of FUV photons in the Doppler core
is not important.
In our calculations, for example, the column density
of H$_2$ molecules easily exceeds 
$N_{\rm H_2} \sim 10^{19}~{\rm cm}^{-2}$, and the FUV photons
are mainly absorbed in the Lorentz wings after that. 
Since the energy range of the Lorentz wings 
is much wider than that of the Doppler core, the Doppler shift
of the core is not so important.

%%%%%%%%%%%%%%%%%%%%%%%%%%%%%%%%%%%%%%%%%%%%%%%%%%%%%%%%%%%%%%%%%%%%%%
%%%%%%                      FIGURES                             %%%%%%
%%%%%%%%%%%%%%%%%%%%%%%%%%%%%%%%%%%%%%%%%%%%%%%%%%%%%%%%%%%%%%%%%%%%%%
%-----------------------------------------------------------------------
\begin{figure}[tb]
\begin{center}
\includegraphics[width=0.5\hsize]{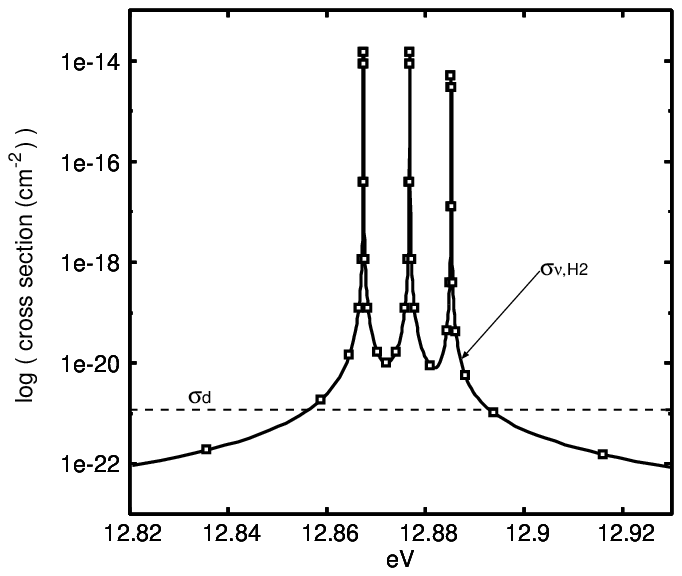}
\caption{Our representative lines for Lyman-bands, 
$\sigma_{\nu, {\rm H_2}}$. The points indicate the center
of the grids. 
The dashed line means the dust absorption cross section
for H$_2$ dissociating FUV photons, $\sigma_{\rm d}$
 }
\label{fig:h2rep}
\includegraphics[width=0.5\hsize]{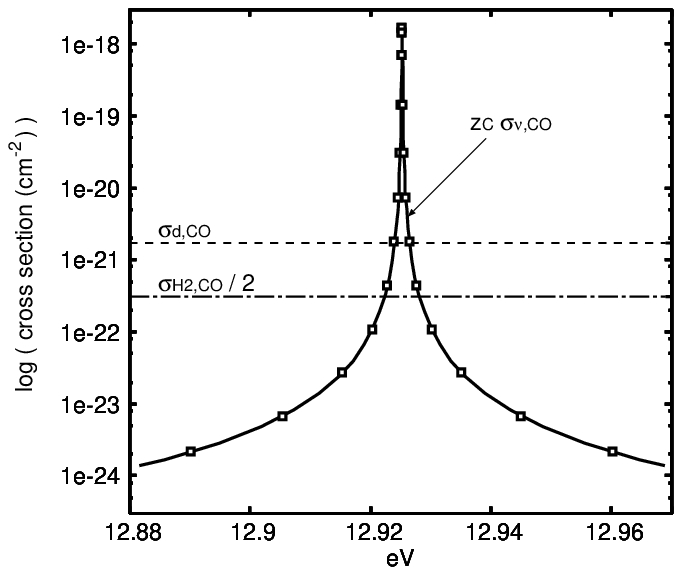}
\caption{Our representative line for the photodissociation
of CO molecules, $Z_{\rm C} \sigma_{\rm \nu, CO}$. 
The points indicate the center of the grids. 
The dashed line (dot-dashed line) means the cross section for
the dust absorption (shielding by H$_2$ molecules),
$\sigma_{\rm d, CO}$ ($\sigma_{\rm H_2, CO}/2$).
 }
\label{fig:corep}
\end{center}
\end{figure}
%-----------------------------------------------------------------------

%%% n dependence %%%
%-----------------------------------------------------------------------
%-----------------------------------------------------------------------
%-----------------------------------------------------------------------
%-----------------------------------------------------------------------
%-----------------------------------------------------------------------
%-----------------------------------------------------------------------
%-----------------------------------------------------------------------
%-----------------------------------------------------------------------
%-----------------------------------------------------------------------
%-----------------------------------------------------------------------
%-----------------------------------------------------------------------
%-----------------------------------------------------------------------

%%% stellar dependence %%%
%-----------------------------------------------------------------------
%-----------------------------------------------------------------------
%-----------------------------------------------------------------------
%-----------------------------------------------------------------------
%-----------------------------------------------------------------------
%-----------------------------------------------------------------------
%-----------------------------------------------------------------------
%-----------------------------------------------------------------------
%-----------------------------------------------------------------------
%-----------------------------------------------------------------------
%-----------------------------------------------------------------------
%-----------------------------------------------------------------------
%-----------------------------------------------------------------------
%-----------------------------------------------------------------------
%-----------------------------------------------------------------------
%------------------------------------------------------------------------

\end{document}